\documentclass[11pt, a4paper]{article}

\usepackage{amsmath,amsfonts,bigints} 
\usepackage{fullpage}
\usepackage{color}
\usepackage{tikz}
\usepackage{tkz-euclide}
\usepackage{url}
\usepackage{authblk}
\usepackage{graphicx} 
\usepackage{hyperref}
\usepackage[left=1.5cm, right=1.5cm, top=1.785cm, bottom=2.0cm]{geometry}

\usepackage{bm}
\usepackage{tikz}
\usepackage{subcaption}
\usepackage{amssymb}
\usepackage{mathtools}
\usepackage{siunitx}
\usepackage{multirow}
\usepackage{array}
\usepackage{setspace}

\definecolor{cream}{RGB}{222,217,201}

\title{Cyclic loading of a heterogeneous non-linear poroelastic material}

\author[1]{Zoe C.~Godard\footnote{Email: {\tt zoe.godard@maths.ox.ac.uk}}}
\author[1]{Derek E.~Moulton}
\author[1]{Sarah L.~Waters}
\affil[1]{Mathematical Institute, University of Oxford, Oxford, UK}
\date{}

\begin{document}

\maketitle

\onehalfspacing

\begin{abstract}
\onehalfspacing
Cyclic loading is a common feature in poroelastic systems, the material response depending non-trivially on the exact form of boundary conditions, pore structure, and mechanical properties. The situation becomes more complex when heterogeneity is introduced in the properties of the poroelastic material, yet heterogeneity too is common in physical poroelastic structures. In this paper, we analyse the behaviour of a soft porous material in response to a uniaxial cyclic stress or displacement, with a focus on understanding how this response is affected by continuous heterogeneity in the stiffness or permeability. Our work is motivated by observed altered material properties of the diseased tendon, but the framework we develop and analyse is generically applicable. We construct a one-dimensional non-linear poroelastic model, assuming Darcy flow through the pores of the solid skeleton which we assume has neo-Hookean elasticity. The system is driven by an applied uniaxial cyclic stress or a uniaxial cyclic displacement at one boundary. Heterogeneity in the stiffness or permeability profile is imposed via a Gaussian bump function. By exploring a range of loading frequencies together with magnitudes and locations of heterogeneity, we characterise the effect of heterogeneity on the response of the material, and show that the response of the system to an applied stress is qualitatively distinct from the response to an applied displacement. Our analysis of this simple model provides a foundation for understanding how heterogeneity affects the poroelastic response to cyclic loading.

\end{abstract}

\newpage

\section{Introduction}

Poroelasticity, though classically derived in the context of soil consolidation \cite{terzaghi:1923,rendulic:1936,biot:1935,biot:1941}, has found applications far beyond this, covering areas as varied as biological tissues, seismology, transport and hydrogeology, to name a few. Their common denominator is the coupling of fluid flow and deformation of a porous solid. Linear poroelasticity theory was first formalised by Biot \cite{biot:1941} and later extended to large deformations \cite{biot:1972}, laying the groundwork still used today for the study of soft porous materials. Detournay and Cheng \cite{detournay:1993} provide a good summary of the origin of poroelasticity and developments following on from Biot's work.

Many poroelastic materials are subject to cyclic loads or deformation throughout their lifetime, ranging from biological to geological systems, with applications in areas such as medicine, bioengineering, construction engineering, climate and drug delivery. For example, soil may experience periodic loading due to passing traffic \cite{jin:2004,hu:2011}, seismic activity (particularly relevant in construction engineering) \cite{genna:1989, popescu:2006, bonazzi:2021}, and waterwaves or tidal forcings on the seabed \cite{yamamoto:1978,madsen:1978,trefry:2019,karim:2002,hu:2011}. Biological soft tissues such as tendon and cartilage also experience periodic tensile and/or compressive loads during daily activities such as walking, swimming, running or playing an instrument \cite{millar:2021,lavagnino:2015,bojsen-moller:2019}. Extensive research supports the hypothesis that tendon (and cartilage) structure is remodelled in response to cyclic mechanical loading \cite{arampatzis:2007,docking:2013,eckstein:1999,lavagnino:2008}, similarly to hard tissue such as bone, where naturally occurring cyclic loading has been shown to play an important role in bone remodelling and healing \cite{piekarski:1977, zhang:1994,manfredini:1999,nguyen:2010,witt:2014}. In addition, there is a rich literature surrounding solute transport in response to cyclic loading of these biological soft tissues, where the effect of parameters such as loading amplitude, frequency, and size of solute have been investigated\cite{zhang:2011,riches:2002,mauck:2003,sengers:2004,cacheux:2023,fiori:2023,fiori:2024,piekarski:1977,witt:2014}. Brain tissue is also commonly modelled as a poroelastic material which may undergo pulsating loads from the dilation of arterioles; in this area there is a particular focus on fluid movement and the clearing of metabolic waste \cite{franceschini:2006,kedarasetti:2022,bojarskaite:2023}.

Although many poroelastic models assume the material to be uniform, poroelastic media commonly exhibit spatially heterogeneous material properties, such as non-uniform stiffness, permeability and Poisson’s ratio. When modelling soil, heterogeneity is typically accounted for by introducing horizontal layers of different thickness, density or permeability, resulting in discrete jumps in the material properties between each layer \cite{genna:1989,popescu:2006,madsen:1978}. Trefry \textit{et al.}\ \cite{trefry:2019} explore the effect of continuous spatial heterogeneity in groundwater systems undergoing cyclic loads (tidal forcings) by imposing a randomly assigned heterogeneous hydraulic conductivity, and find this leads to chaotic mixing. Concerning biological tissues, Zhang \cite{zhang:2011} investigates the effect of linear variation of stiffness and permeability, chosen to fit experimental data for cartilage depth-dependent material properties, in a cylindrical cartilage disc subject to cyclic compression. Similarly to the layered soil models, Kameo \textit{et al.}\ \cite{kameo:2016} and Ruiz, \textit{et al.}\ \cite{ruiz:2013} assign unique material properties to vertical layers of trabecula (anatomical unit of bone) and intervertebral disc (IVD) respectively. While these models all find that the various forms of heterogeneity significantly impact the response of the material, to date there is no systematic study investigating the effect of generic continuous material heterogeneity on the poroelastic response to cyclic loading. Additionally, the biological models consider material heterogeneity to be a characteristic of the healthy tissue, however in tissues such as tendon several studies have reported altered mechanical and material properties in the damaged tendon compared to its healthy counterpart\cite{arya:2010,obst:2018}. Damaged tissue has primarily been explored as cracks or fractures in the material\cite{nguyen:2011,orozco:2022}, as in rock mechanics \cite{ambartsumyan:2019}, rather than modified material properties. In this study, we will refer to “damage” as heterogeneity in the material properties of the tissue.

Our study is particularly motivated by tendinopathy, a common and painful condition affecting tendons and characterised by a modified structure of the tissue \cite{millar:2021}. Modelling the mechanics of such a complex structure forms a formidable challenge, for which a number of distinct approaches exist. Tendons have been modelled as purely elastic (these models are usually fibre reinforced) \cite{bajuri:2016,shearer:2015,shearer:2015a}, viscoelastic \cite{gupta:2010}, poroelastic \cite{wren:2000,safa:2020,lavagnino:2008,atkinson:1997,butler:1997,chen:1998}, or even poroviscoelastic \cite{khayyeri:2015}. Some models have also addressed the evolution of material properties in tendon. For example Wren \textit{et al.}\ \cite{wren:2000} incorporated a phenomenological rule to update global tissue permeability, and Lavagnino \textit{et al.}\ \cite{lavagnino:2008} built a submodel for the tendon cell to explore the relation between physical stimuli and protein production (mechanotransduction). A number of these models are finite element models; while they can enable additional levels of biological detail to be captured \cite{safa:2020,lavagnino:2008,chen:1998,atkinson:1997,butler:1997,khayyeri:2015,bajuri:2016}, this additional complexity can cloud identification of the underlying driving physical mechanisms governing the response of the system. Tendons are able to withstand large strains \cite{thorpe:2015}, motivating a non-linear framework; a one-dimensional formulation, as we will present here, allows for non-linear kinematics whilst remaining simple enough to draw physical and intuitive understanding of the underlying mechanics.

In this paper, we present a non-linear poroelastic model for a heterogeneous material subject to a uniaxial tensile cyclic stress or displacement of varying frequency. Of particular relevance to our study is the work of Fiori \textit{et al.}\ \cite{fiori:2023}, who conducted a systematic study exploring the effect of compressive cyclic applied displacement amplitude and frequency on the fluid response of a homogeneous poroelastic material in a uniaxial setting. On the one hand, they found that as the period decreases and the amplitude increases the root-mean-square relative error between the linear and non-linear model grows, identifying the validity of the linear poroelastic model as $A \ll \sqrt{T}$, where $A$ and $T$ are the dimensionless amplitude and period of applied displacement, such that even at small amplitude $A$ linear poroelasticity does not provide a good approximation for high frequency loading (small $T$). In addition, the authors found that decreasing loading period increases localisation of the deformation while varying loading amplitude and initial porosity has little qualitative effect on the poroelastic response. The results from Fiori's study for the fluid response of a homogeneous poroelastic material to cyclic compression qualitatively agree with the results from the uniform version of our model, despite employing a different solid constitutive law, though we might expect this choice to play a greater role when heterogeneity is present. The aim of this paper is to systematically investigate and intuitively understand the role of material heterogeneity on the response of a poroelastic material to cyclic loading. To this end, we explore the effect of different forms of heterogeneity, located at different positions in the material. We also analyse how the consequences of heterogeneity vary with loading frequency, and we uncover both quantitative and qualitative differences in the material response to an applied displacement versus an applied stress. 

Following Macminn \textit{et al.}\ \cite{macminn:2016} and Coussy\cite{coussy:2004}, we introduce a non-linear poroelastic model which in its reference state has uniform porosity and position-dependent permeability, and non-linear elasticity with position-dependent stiffness. We employ a Lagrangian framework that better facilitates incorporation of material heterogeneity. Assuming uniaxial deformation, we apply a tensile cyclic stress or displacement and fixed pressure at one end with no flow and displacement through the other end, and solve the full problem numerically using a finite volume method and stiff ode solver. We first impose damage as a local decrease in stiffness, and investigate the effect of this heterogeneity on the strain and flux response of the material for a fixed loading frequency and amplitude. We then characterise how the response varies with the magnitude and location of the decrease, as well as with loading frequency. We also consider the effect of a local decrease in permeability and compare to our results for heterogeneous stiffness. Finally, we discuss the physical interpretation of the model, including relevance of our findings to tendon and how our mechanistic insights might contribute to our understanding of tendinopathy development or diagnosis.

\section{Model}

We consider a fully saturated poroelastic material composed of a solid and a fluid phase, such that $\phi_f + \phi_s = 1$ where $\phi_{f,s}$ are the true fluid and solid volume fractions respectively. Here the word “true” is used interchangeably with “Eulerian”, meaning a quantity measured with respect to the current configuration. This is contrasted with “nominal” or “Lagrangian” quantities, which are measured with respect to the reference (initial) configuration. For example, the true porosity $\phi_f$ measures the current fluid volume fraction of the total current volume $dv$, while the nominal porosity $\Phi$ measures the current fluid volume fraction of the total reference volume $dV_0$, where capital letters are used to denote Lagrangian variables. The true and nominal porosities are related by $\Phi_{f,s}=J\phi_{f,s}$ where $J$ is the Jacobian determinant which measures local volume change i.e.\ $dv=JdV_0$, and is defined later in equation \eqref{Eq.Jacobian}. For the fully saturated poroelastic material in a Lagrangian description, we have $\Phi_f + \Phi_s = J$. Going forward, we drop subscripts and use only $\Phi\equiv\Phi_f$ ($\phi\equiv\phi_f$) to denote nominal (true) porosity. Supposing that the material begins at time $t=0$ in a stress free state, the reference or initial configuration refers to the material's configuration at time $t=0$, while the current configuration refers to the material's configuration at current time $t$.

Since the fluid flow and solid deformation are coupled, we are presented with a choice of framework: Lagrangian or Eulerian. Most poroelastic descriptions are presented in an Eulerian frame, as this is most natural for fluid flow, but it is more natural to consider material heterogeneity in a Lagrangian frame, as this gives immediate access to material points and their properties. While we restrict to fixed heterogeneity in this paper, a Lagrangian frame could also more readily accommodate complexities such as tissue remodelling, via an evolution of properties at material points. 

\begin{figure}[h]
    \centering
    \includegraphics[width=9cm]{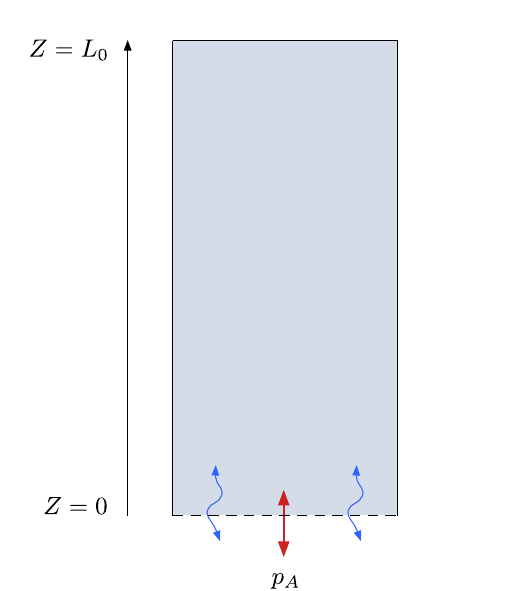}
    \caption{We consider a poroelastic material of length $L_0$ which is pulled at $Z=0$ (red arrow) where it is also subject to ambient fluid pressure $p_A$, and fixed at $Z=L_0$ with no deformation or flow. The blue arrows represent fluid flow.  The material is shown in the reference configuration.}
    \label{Fig. Diagram}
\end{figure}

We therefore consider the uniaxial deformation of a poroelastic material in the Lagrangian domain
\begin{equation}0\leq Z\leq L_0\end{equation}
where $L_0$ is the length of the material, as shown in Figure \ref{Fig. Diagram}. We assume the boundaries of the system coincide with the boundaries of the solid. The material is pulled at $Z=0$ where it is also subject to ambient fluid pressure $p_A$. The upper boundary $Z=L_0$ is fixed with no deformation or flow.

\subsection{Kinematics}

The deformation gradient tensor $\mathbf{F}$ is given by 
\begin{equation}
    F_{ij} = x_{i,j} = \frac{\partial x_i}{\partial X_j}.
\end{equation}
where $\mathbf{X}$ and $\mathbf{x}$ denote Lagrangian and Eulerian coordinates respectively, and $x_i$ and $X_j$ are components with $i,j=1,2,3$. The Jacobian determinant is defined by 
\begin{equation}\label{Eq.Jacobian}
    J=\det(\mathbf{F}).
\end{equation} 
For uniaxial deformation, the deformation gradient tensor is then given by
\begin{equation}\label{Eq. F 1D}
    \textbf{F} = \begin{pmatrix}
        1&0&0\\
        0&1&0 \\
        0&0&J
    \end{pmatrix}.
\end{equation}
The solid displacement $\mathbf{U}$ is defined as the difference between the current configuration $\mathbf{x}(\mathbf{X},t)$ and the reference (initial) configuration $\mathbf{X}$ of the solid:
\begin{equation}
    \mathbf{U} = \mathbf{x}(\mathbf{X},t)-\mathbf{X},
\end{equation}
where $\mathbf{x}(\mathbf{X},t)$ maps the reference state to the current state. The material begins at $t=0$ in the reference state, i.e.\ $\mathbf{U}(\mathbf{X},0) = \mathbf{0}$. 
The deformation gradient tensor can then be expressed as
\begin{equation}\label{Eq. F wrt displacement}
    \textbf{F} = \nabla_X(\mathbf{x}) = \textbf{I} + \nabla_X(\mathbf{U})
\end{equation}
where $\nabla_X$ denotes the Lagrangian gradient operator in Lagrangian coordinates ($\partial/\partial X_I$) and $\textbf{I}$ is the identity tensor. For future use, $\nabla_x$ denotes the Eulerian gradient operator ($\partial/\partial x_i$). For uniaxial deformation then $\mathbf{U} = (0,0,U)$ so that strain
\begin{equation}\label{eq: U_Z = J}
    J = 1+ U_Z
\end{equation}
where $\mathbf{X} = (X,Y,Z)$ and we have used subscript $Z$ as a shorthand for $\partial / \partial Z$.

\subsection{Continuity}

We next derive local conservation equations to describe the deformation. It is more natural to write these first in terms of Eulerian quantities, which we then convert into a Lagrangian frame. We consider both the solid grains which form the solid skeleton and the fluid to be incompressible, so that when the poroelastic material is compressed the pore size must decrease (fluid is squeezed out), i.e.\ the solid fraction increases as the total volume decreases and vice versa. The fully saturated poroelastic material is therefore compressible through pore rearrangement. Mathematically, the incompressibility condition for the solid phase is expressed as $(1-\phi)dv = (1-\phi_0)dV_0$ or $(J-\Phi)dV_0 = (1-\Phi_0)dV_0$ where $\phi_0 = \Phi_0$ is the initial porosity. The change in volume is then
\begin{equation}\label{Eq.Jacphi}
    J = \frac{1-\phi_0}{1-\phi} = 1 + \Phi - \Phi_0 \equiv 1+\hat{\Phi},
\end{equation}
where we have defined the normalised quantity $\hat{\Phi}\equiv\Phi-\Phi_0$. Equation \eqref{eq: U_Z = J} then becomes:
\begin{equation}\label{Eq. Uz=Phi)}
    U_Z = \hat{\Phi}.
\end{equation}

Since the fluid and solid are incompressible, and we assume uniform density for both phases, the fluid and solid mass densities are constant. Local continuity for the solid and fluid phases is then written
\begin{subequations}\label{Eq. Continuity 1}
    \begin{equation}\label{Eq. fluid cont 1}
        \frac{D^f}{Dt}\int_{v(t)}\phi dv = 0
    \end{equation}
    \begin{equation}\label{Eq. solid cont 1}
        \frac{D^s}{Dt}\int_{v(t)}(1-\phi) dv = 0,
    \end{equation}
\end{subequations}
where $v(t)$ is an arbitrary current volume, and $D^f/Dt$ and $D^s/Dt$ are respectively the material derivatives with respect to the fluid and solid (i.e.\ the time derivative that an observer attached to a fluid or a solid particle would derive \cite{coussy:2004}). 

Let $\mathbf{v}^{s,f}\equiv \mathbf{V}^{s,f}$ be the solid and fluid velocities respectively. Note that although we write velocities (and subsequently pressure) in lowercase for Eulerian and uppercase for Lagrangian, they are the same in both of these frames -- only the coordinate at which they are evaluated changes. The continuity equations \eqref{Eq. Continuity 1} can be rewritten in terms of arbitrary reference volume $V_0$ and using Euler's identity:
\footnote[5]{$\frac{D^{f,s}J}{Dt} = J\nabla_x\cdot\mathbf{v}^{f,s}$} 
\begin{subequations}\label{Eq. Continuity 2}
    \begin{equation}\label{Eq. fluid cont 2}
        \frac{D^s \phi}{Dt}+\nabla_x\cdot\mathbf{q}+\phi\nabla_x\cdot\mathbf{v}^s = 0
    \end{equation}
    \begin{equation}\label{Eq. solid cont 2}
         \frac{D^s \phi}{Dt}-(1-\phi)\nabla_x\cdot\mathbf{v}^s = 0,
    \end{equation}
\end{subequations}
where we have also defined the Eulerian relative flow $\mathbf{q} = \phi(\mathbf{v}^f-\mathbf{v}^s)$. From Nanson's formula (see appendix \ref{app: Surf trans}) the Lagrangian relative flow $\mathbf{Q}$ is equal to Eulerian relative flow:
\begin{equation}\label{Eq: Q Vs}
    \mathbf{Q} = \frac{\Phi}{J}(\mathbf{V}^f-\mathbf{V}^s),
\end{equation}
where we used $\Phi = J\phi$. Equations \eqref{Eq. Continuity 2} are converted into Lagrangian coordinates using the deformation gradient tensor \eqref{Eq. F 1D} to arrive at the Lagrangian equations for fluid conservation of mass as follows:
\begin{subequations}\label{Eq. non-linear flow equation 3D}
    \begin{equation}\label{Eq. fluid cont FINAL 3D}
    \frac{d\Phi}{dt}+\nabla_X\cdot\mathbf{Q} = 0
    \end{equation}
    \begin{equation}\label{Eq. solid cont FINAL 3D}
        \nabla_X\cdot\mathbf{Q} + J(\mathbf{F}^{-1}\nabla_X)\cdot\mathbf{V}^s = 0
    \end{equation}
\end{subequations}
where we have denoted $D^s/Dt \equiv d/d t$. For uniaxial deformation we have $\mathbf{V}^{s,f} = (0,0,V^{s,f})$ and $\mathbf{Q} = (0,0,Q)$, and the continuity equations \eqref{Eq. non-linear flow equation 3D} simplify to:
\begin{subequations}\label{Eq. non-linear flow equation}
    \begin{equation}\label{Eq. fluid cont FINAL}
    \frac{d \Phi}{d t} + \frac{\partial Q}{\partial Z} = 0
    \end{equation}
    \begin{equation}\label{Eq. solid cont FINAL}
        \frac{\partial Q}{\partial Z} + \frac{\partial V^s}{\partial Z} = 0.
    \end{equation}
\end{subequations}
Note that the solid continuity equation \eqref{Eq. solid cont FINAL} can also be replaced by Eq.\eqref{Eq.Jacphi}, as the incompressibility condition automatically satisfies conservation of solid mass. The continuity formulation \eqref{Eq. solid cont FINAL}, however, will be useful when implementing boundary conditions.

\subsection{Constitutive law for fluid flow}

Let $p(\mathbf{x},t) \equiv P(\mathbf{X},t)$ be the fluid pore pressure. We assume Darcy's law to govern the relative flow of the fluid with respect to the solid, neglecting gravity. In an Eulerian frame it is given by
\begin{equation}\label{Eq. Eulerian Darcy}
    \mathbf{q} = -\frac{\mathbf{k}}{\mu}\nabla_x p,
\end{equation}
where $\mathbf{k}$ is the permeability tensor and $\mu$ is dynamic viscosity. The Lagrangian flux $\mathbf{Q}$ is then
\begin{equation}\label{Eq: Q dP 3D}
    \mathbf{Q} = -J\frac{\mathbf{k}}{\mu}\mathbf{F}^{-1}\mathbf{F}^{-T}\nabla_X P.
\end{equation}
Letting $k$ denote permeability in the $Z$-direction, then under uniaxial deformation the flux is given by
\begin{equation}\label{Eq: Q dP}
    Q = -\frac{k}{\mu}\frac{1}{J}\frac{\partial P}{\partial Z}.
\end{equation}
We assume permeability is porosity dependent and adopt the normalised Kozeny-Carman formula \cite{macminn:2016}, widely used for a range of materials. This formula reads
\begin{equation}\label{Eq. KC}
    k_{KC}(\phi) = k_0\frac{(1-\phi_0)^2}{\phi_0^3}\frac{\phi^3}{(1-\phi)^2},
\end{equation}
which we express in an Eulerian frame since it is physically more meaningful, and $k_0$ is the reference permeability value. In particular, it captures two important physical limits: as the true porosity vanishes ($\phi\to 0$), permeability vanishes, and as the true porosity tends to 1 ($\phi\to 1$), permeability diverges. This prevents the flow from driving true porosity below zero and above 1. An important consequence of this law is that permeability increases with porosity, which is consistent with observations in tendons \cite{chen:1998,kent:2023}. In Lagrangian, and with the expression \eqref{Eq.Jacphi} for $J$, the formula reads
\begin{equation}\label{Eq. KC Lag}
    k_{KC}(\Phi) = k_0\frac{(\Phi/\Phi_0)^3}{1+\Phi - \Phi_0}.
\end{equation}

\subsection{Mechanical equilibrium}

Following Macminn \cite{macminn:2016}, the true total stress $\bm{\sigma}$, which describes the current total force supported by the fluid and solid per current unit total area,  can be divided into the true fluid stress $\bm{\sigma}^s$ and true solid stress $\bm{\sigma}^f$ (fluid and solid force per unit fluid and solid area respectively), assuming the phase area and phase volume fractions are equivalent: $\bm{\sigma} = (1-\phi)\bm{\sigma}^s+\phi\bm{\sigma}^f$. We can write $\bm{\sigma}^f = -(p-p_A)\mathbf{I}$ \cite{macminn:2016,coussy:2004}. Since the fluid is coupled to the solid skeleton, the fluid exerts at all times a pressure on the solid, so that the total solid stress includes this component even when the fluid is at rest. This component of the stress cannot contribute to the deformation of the solid skeleton of incompressible grains, motivating the use of Terzaghi's effective stress $\bm{\sigma'} \equiv (1-\phi)(\bm{\sigma}^s + (p-p_A)\textbf{I})$. The true total stress is then
\begin{equation}
    \bm{\sigma} = \bm{\sigma'}-(p-p_A)\textbf{I}.
\end{equation}
To satisfy mechanical equilibrium, the solid skeleton and the fluid must jointly support the mechanical stress. In the absence of body forces and inertia, this requires
\begin{equation}\label{Eq.sigeq}
    \nabla_x \cdot \bm{\sigma} = \nabla_x\cdot \bm{\sigma'} - \nabla_xp = \mathbf{0}
\end{equation}
in an Eulerian frame. To describe mechanical equilibrium in the Lagrangian solid frame, let $\mathbf{s}$ be the nominal total stress, also known as the first Piola-Kirchoff stress tensor. It is related to the Cauchy stress tensor $\bm{\sigma}$ by 
\begin{equation}\label{Eq.sig2s}
    \mathbf{s} = J\bm{\sigma}\textbf{F}^{-T},
\end{equation}
and the effective nominal stress $\mathbf{s'}$ is then $\mathbf{s} = \mathbf{s'}-J\textbf{F}^{-T}p$. Combining \eqref{Eq.sigeq} and \eqref{Eq.sig2s}, we arrive at the following equation for mechanical equilibrium in the Lagrangian frame:
\begin{equation}\label{Eq.MechEq total}
    \nabla_X(\mathbf{s'}) = \nabla_X(J\mathbf{F}^{-T}P).
\end{equation}
Under uniaxial deformation this may further be simplified to
\begin{equation}\label{Eq.MechEq}
    \frac{\partial s'}{\partial Z} = \frac{\partial P}{\partial Z},
\end{equation}
where $s'$ denotes the $ZZ$-component of the stress tensor.

\subsection{Constitutive law for the solid skeleton}

Incompressibility of the solid grains means that the volume of solid contained within the Lagrangian element is fixed and cannot change, however the current volume of the material element can change freely through fluid flow and pore rearrangement. The constitutive law for the effective solid stress describes the “drained” behaviour of the solid skeleton - this corresponds to what would be measured for the “dry” skeleton, or if the “wet” skeleton were tested sufficiently slowly so that the fluid is stationary. Since the network forming the solid skeleton is compressible, a compressible constitutive law must be considered. Some care must be taken in choosing an appropriate effective stress law for a poroelastic solid, as highlighted by Zheng and Zhang \cite{zheng:2019}, who find that only the Green-strain tensor is compatible with the restrictions of thermodynamics. Accordingly, we consider a neo-Hookean model for a compressible solid:
\begin{equation}\label{neo-hookean}
    \mathbf{s'} = \frac{\partial W}{\partial \textbf{F}}, \quad W = \frac{M-\Lambda}{4}(1+J^2-2\ln{J})+\frac{\Lambda}{2}(J-1)^2
\end{equation}
where $W$ is the strain energy density, $M$ is the oedometric stiffness and $\Lambda$ is Lamé's second parameter, related to $M$ by Poisson's ratio $\nu$
\begin{equation}\label{Eq. Lambda}
    \Lambda = \frac{\nu}{1-\nu}M.
\end{equation}
As a form of incorporating material heterogeneity, we consider the general case where $M$ and $\Lambda$ may be functions of position. The $ZZ$-component of effective stress is thus given by
\begin{equation}\label{Eq. stress law}
    s'(J) = \frac{M(Z)}{2}\left(J-\frac{1}{J}\right)+\frac{\Lambda(Z)}{2}\left(J+\frac{1}{J}-2\right).
\end{equation}
Note that for $\nu = 0.5$ we have $s' = M(J-1) = MU_Z$, which is linear.

Eqs. \eqref{Eq.Jacphi}, \eqref{Eq. fluid cont FINAL}, \eqref{Eq: Q dP}, \eqref{Eq. KC Lag}, \eqref{Eq.MechEq} and \eqref{Eq. stress law} with specified boundary and initial conditions, may be combined to form a closed model for the evolution of porosity $\Phi$ in a Lagrangian framework, as will be highlighted in the summary \ref{subsec: summary}.

\subsection{Boundary and initial conditions}

\subsubsection{Lower boundary \texorpdfstring{$Z=0$}{Z=0}.}
We subject the poroelastic material to cyclic tension via applying either a periodic stress or periodic displacement boundary condition at $Z=0$. Adopting the convention that tension is positive and compression is negative, the tensile applied stress (AS) takes the form
\begin{subequations}
\begin{equation}
    s^*(t) = A_l[1-\cos(\omega t)],
\end{equation}
while the tensile applied displacement (AD) is given by
\begin{equation}
    \quad a(t) = -\frac{A_d}{2}[1-\cos(\omega t)].
\end{equation}
\end{subequations}
Here $A_{l,d}$ are the magnitudes of the applied stress and applied displacement, respectively, and $\omega$ is the frequency. Both scenarios generate tensile stress, since $s^*(t)\geq0$ and $a(t)\leq0$ for all time. Note that these conditions are not equivalent: applying a stress with amplitude $A_l$ does not result in a displacement of amplitude $A_l$ at the boundary, and vice versa. We also define the loading period $T=2\pi/\omega$. The associated boundary conditions are
\begin{subequations}\label{Eq: BCs 1}
\begin{equation}
    s(Z=0,t) = s^*(t)
    \label{eq: BC stress}
\end{equation}
\begin{equation}
        \text{or}\quad U(Z=0,t) = a(t).
    \label{eq: BC disp}
\end{equation}
\end{subequations}
For an applied displacement the boundary condition can also be expressed as $V^s(Z=0,t)=da/dt\equiv\dot{a}$ (overdot denotes derivative with respect to time). It will be useful to differentiate between loading time periods when the material is being “pulled”, characterised by $\dot{s}^*>0$ or $\Dot{a}<0$, versus the unloading or relaxing time periods when the material is “let go” or “pushed” back, for which $\dot{s}^*<0$ or $\Dot{a}>0$. 

In terms of the fluid, we suppose that the $Z=0$ boundary is permeable and held at a fixed pressure, that is
\begin{equation}
    p(Z=0,t) = p_A.
\end{equation}

\subsubsection{Upper boundary \texorpdfstring{$Z=L_0$}{Z=L_0}.}

The $Z=L_0$ boundary is assumed fixed and impermeable to fluid, so that:
\begin{equation}\label{Eq. BC Z=1}
   U(Z=L_0,t)=0,\quad V^f(Z=L_0,t)=0.
\end{equation}
Note that this also implies that $V^s(Z=L_0,t)=0$ and so $Q(L_0,t) = 0$. Combined with solid continuity \eqref{Eq. solid cont FINAL} we find that
\begin{equation}\label{eq: Vs = -Q}
    V^s = -Q
\end{equation}
everywhere, and from \eqref{Eq: Q Vs}
\begin{equation}
    V^f =- V^s\biggl(\frac{J-\Phi}{\Phi}\biggl),
\end{equation}
so that the fluid and the solid always locally move in opposite directions, since $J-\Phi = 1-\Phi_0 > 0$ from \eqref{Eq.Jacphi}.

\subsubsection{Initial condition.}
The material starts in the reference configuration, corresponding to a relaxed or undeformed state, with initial uniform porosity $\Phi_0$
\begin{equation}\label{Eq. IC}
    U(Z,0) = 0 \quad\text{and}\quad \Phi(Z,t) = \Phi_0.
\end{equation}

\subsection{Non-dimensionalisation}

We non-dimensionalise as follows:
\begin{gather}
    Z = L_0\tilde{Z} \quad U = L_0\tilde{U}^s \quad t = T_{pe}\tilde{t} \quad s = M_0\tilde{s} \nonumber \\
    \omega = \frac{\tilde{\omega}}{T_{pe}} \quad A_l = M_0\tilde{A_l} \quad A_d = L_0\tilde{A_d} \nonumber \\
    k = k_0\tilde{k} \quad P = M_0\tilde{P} + p_A \quad M = M_0\tilde{M} \quad \Lambda = M_0\frac{\nu}{1-\nu}\tilde{M}
\end{gather} 
where $k_0$ and $M_0$ are the scalar reference permeability and oedometric stiffness and $T_{pe} = \mu L_0^2/k_0M_0$ is the classical poroelastic timescale which corresponds to the characteristic diffusion timescale of the material, i.e.\ the time it takes for the response to an applied stress/displacement  at $Z=0$ to reach $Z=L_0$. The stiffer, shorter and more permeable the material, the quicker the response diffuses through the material. For a constant applied stress or displacement the poroelastic timescale corresponds to time it takes for the system to reach steady state. For a cyclic applied stress/displacement, the loading frequency $\tilde{\omega}$ falls into two different regimes: $\tilde{\omega} \leq 2\pi$, for which the diffusion time is less than or equal to the loading period, and $\tilde{\omega}>2\pi$, for which the diffusion time is greater than the loading period. The smaller $\tilde{\omega}$ is, the more time the stress imposed at $Z=0$ has to diffuse through the material, homogenising the response. On the other hand, the greater $\tilde{\omega}$ is, the more localised the response becomes to $Z=0$. Note that varying $\tilde{\omega}$ may correspond to varying $\omega$, $L_0$, $k_0/\mu$ or $M_0$. Going forward we consider the non-dimensional problem and drop the tilde notation, unless specified otherwise.

\subsection{Parameter values}\label{subsec: parameter values}

Our choice of parameter values is motivated by tendon, however these can easily be modified for other materials or tissues. Reported values for tendon are variable, with particular disagreement on the Poisson's ratio \cite{gatt:2015,reese:2010,thorpe:2014}. We choose the following \cite{lavagnino:2008,yoo:2007,kurihara:2012}:
\begin{equation}
\begin{split}
     \nu = 0.3 \quad E_0 = 1\unit{GPa} \quad L_0 = 3\unit{cm} \\
     \frac{k_0}{\mu} = 3.98 e^{-14} \unit{m} \unit{N^{-1}s^{-1}} \quad \Phi_0 = 0.55
\end{split}
\end{equation}
where $M_0$ is calculated from Young's modulus $E_0$ using
\begin{equation}
    M_0 = \frac{E(1-\nu)}{(1+\nu)(1-2\nu)}.
\end{equation}

Tendons, in particular energy-storing tendons, undergo large deformation, with strain ranging from $5\%$ to as large as $30\%$ \cite{thorpe:2015}.
As we will discuss later, the system is diffusive so that for homogeneous stiffness strain $U_Z$ decreases with $Z$: to bound strain throughout the material we simply need to bound strain at $Z=0$.
Setting $0.05\leq U_Z(0,t) \leq 0.3$ and recalling that $J = 1+U_Z$, we calculate the corresponding neo-Hookean stress using Eq. \eqref{Eq. stress law} with the above parameters (and uniform stiffness $M=1$) to find an approximate range for $A_l$: $0.05 \leq A_l\leq 0.3$. Unsurprisingly, $A_l \approx U_Z(0,t)$, since for $\nu=0.3$ we have $s' \approx MU_Z$ for $U_Z<0.5$. For an applied displacement, we look for a range for $U(0,t) = a(t)$. Since strain decreases with $Z$ we have $U(0,t) = \int_1^0U_ZdZ < U_Z(0,t)$. Accordingly, the range for $U(0,t)$ is smaller than the range for $U_Z(0,t)$ so we set $0.05\leq A_d\leq 0.2$. Fiori \textit{et al.}\ \cite{fiori:2023} examined the impact of loading amplitude for a material subject to cyclic compression and demonstrated it had little to no qualitative effect on the poroelastic response. Accordingly, we do not expect loading amplitude to qualitatively affect the poroelastic response of the material subject to cyclic tension. In this study, we will therefore vary loading frequency and fix loading amplitude:
\begin{equation}
    A_l = 0.2 \quad \text{and} \quad A_d = 0.1.
\end{equation}
While the quantitative results will be tied to these specific values of amplitude, we have confirmed that the results presented are qualitatively similar at the lower end of our amplitude (results not shown).

\subsection{Material heterogeneity} 

\begin{figure}[h!]
    \centering
    \includegraphics[width=9cm]{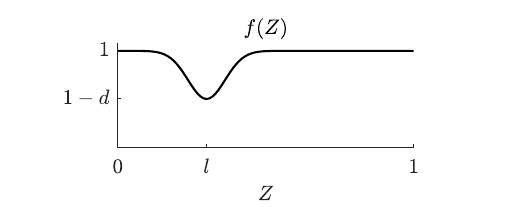}
    \caption{Material damage is modelled with an inverted Gaussian $f$ of magnitude $d$ at $Z=l$}
    \label{Fig: Gaussian}
\end{figure}

Our primary interest in this paper is the effect of material heterogeneity on the response to cyclic stress or displacement driven loading. Motivated by observed material changes in damaged biological tissues, where for instance diseased tendon exhibits a localised decrease in stiffness, we consider localised decreases in either stiffness or permeability.  We impose heterogeneity in stiffness or permeability via a function $f(Z)$ such that
\begin{equation}\label{Eq. Damaged material property}
    M(Z) = f(Z) \quad \text{and} \quad k(\Phi,Z) = k_{KC}(\Phi)f(Z).
\end{equation}
We consider the following form for $f$:
\begin{equation}\label{Eq: Gaussian}
     f(Z) = 1- d\exp\biggl(-\frac{(Z-l)^2}{2c^2}\biggl)
\end{equation}
where $l$ and $d$ are the location and magnitude of damage such that $f(l) = 1-d$, as shown in Figure \ref{Fig: Gaussian}. We set the variance to $c=1/10$. Conversely, an increase may also be considered by taking $2-f(Z)$.  Although our analysis will be focused on this form for $f$, the framework can easily accommodate for other functional forms of heterogeneity.

\subsection{Summary}\label{subsec: summary}

Expressed in terms of $\hat{\Phi}$, the system evolves according to a non-linear advection-diffusion equation,
\begin{equation}\label{Eq:1D Non-linear flow Summary}
     \frac{\partial \hat{\Phi}}{\partial t} = \frac{\partial}{\partial Z}\biggl[A(\hat{\Phi}) + D(\hat{\Phi}) \frac{\partial\hat{\Phi}}{\partial Z}\biggl],
\end{equation}
where
\begin{equation}\label{Eq. Advection Diffusion Summary}
    A(\hat{\Phi}) \equiv \frac{k}{1+\hat{\Phi}}\frac{dM}{dZ}\frac{\partial s'}{\partial M} \quad \text{and} \quad D(\hat{\Phi}) \equiv \frac{k}{1+\hat{\Phi}}\frac{\partial s'}{\partial\hat{\Phi}},
\end{equation}
which respectively represent the bulk advection and bulk diffusivity of the material. The advection term is only present when stiffness $M$ is heterogeneous. The effective stress is given by
\begin{equation}\label{Eq. stress law Summary}
    s' = \frac{M}{2}\left(J-\frac{1}{J}\right)+\frac{\Lambda}{2}\left(J+\frac{1}{J}-2\right), \quad J=1+\hat{\Phi},
\end{equation}
where stiffness for the homogeneous and heterogeneous cases is respectively given by
\begin{equation}\label{Eq: stiffness Summary}
    \begin{split}
    M=1 \quad &\text{and}\quad \Lambda = \frac{\nu}{1-\nu}\\
    \text{or}\quad M=f(Z) \quad &\text{and}\quad \Lambda = \frac{\nu}{1-\nu}f(Z),    
\end{split}
\end{equation}
and similarly permeability is given by
\begin{equation}\label{Eq: perm Summary}
    k = k_{KC} \quad \text{or} \quad k = k_{KC}f(Z)
\end{equation}
with permeability law
\begin{equation}\label{Eq. perm law Summary}
    k_{KC}(\hat{\Phi}) = \frac{(\hat{\Phi}/\Phi_0+1)^3}{1+\hat{\Phi}}.
\end{equation}
The boundary condition on $\hat{\Phi}$ at $Z=0$ is
\begin{equation}\label{Eq. BC Z=0 Summary}
         \quad \hat{\Phi} = \Phi^*(t) \quad \text{or} \quad -V^s = A(\hat{\Phi}) + D(\hat{\Phi})\frac{\partial \hat{\Phi}}{\partial Z} = -\Dot{a} 
\end{equation}
where $\Phi^*$ is calculated by inverting the relationship $s'(\hat{\Phi})$ for given $s^*$. The boundary condition on $\hat{\Phi}$ at $Z=1$ is
\begin{equation}\label{Eq. BC Z=1 Summary}
        Q = A(\hat{\Phi}) + D(\hat{\Phi})\frac{\partial \hat{\Phi}}{\partial Z} = 0.
\end{equation}
Finally, we impose the initial condition
\begin{equation}\label{Eq: IC Summary}
    \hat{\Phi}(Z,0)=0.
\end{equation}

As discussed, working in a Lagrangian framework (with respect to the solid) is the natural choice when considering material heterogeneity. This also presents a significant advantage when solving the system, as the boundaries remain stationary so that the equations can be solved on a fixed domain, unlike in an Eulerian framework which would involve a moving boundary. We solve the system \eqref{Eq:1D Non-linear flow Summary}-\eqref{Eq: IC Summary} numerically for $\hat{\Phi}$ using a finite volume method in space and an implicit variable-step size method in time via the function \texttt{ode15s} in MATLAB (see Appendix \ref{App: method} for details).

\section{Results}

We first consider heterogeneous stiffness for a fixed loading frequency $\omega$ and amplitude $A_{l,d}$. We compare the response of the system for a fixed damage magnitude and three damage locations to the response of the undamaged system over one steady cycle. By steady cycle we mean the code was run for sufficiently long so that the transient behaviour decays (and the response has settled down to a periodic state). We next define a metric to capture the effect of damage, allowing us to explore a range of damage magnitudes, damage locations and loading frequencies. We also briefly explore the effect of heterogeneous permeability in comparison to heterogeneous stiffness. Note that we will refer to normalised porosity as porosity for brevity.

\subsection{Heterogeneous stiffness}\label{Sec: Het stiff}

Consider heterogeneous stiffness $M = f(Z)$ and $\Lambda = \nu f(Z)/(1-\nu)$. We only consider a local decrease in stiffness here, however we also display results for a local increase in stiffness in Appendix \ref{App. increased stiffness}. 

\subsubsection{Full steady response}

\begin{figure}[h!]
\centering
\includegraphics[width=14.9cm]{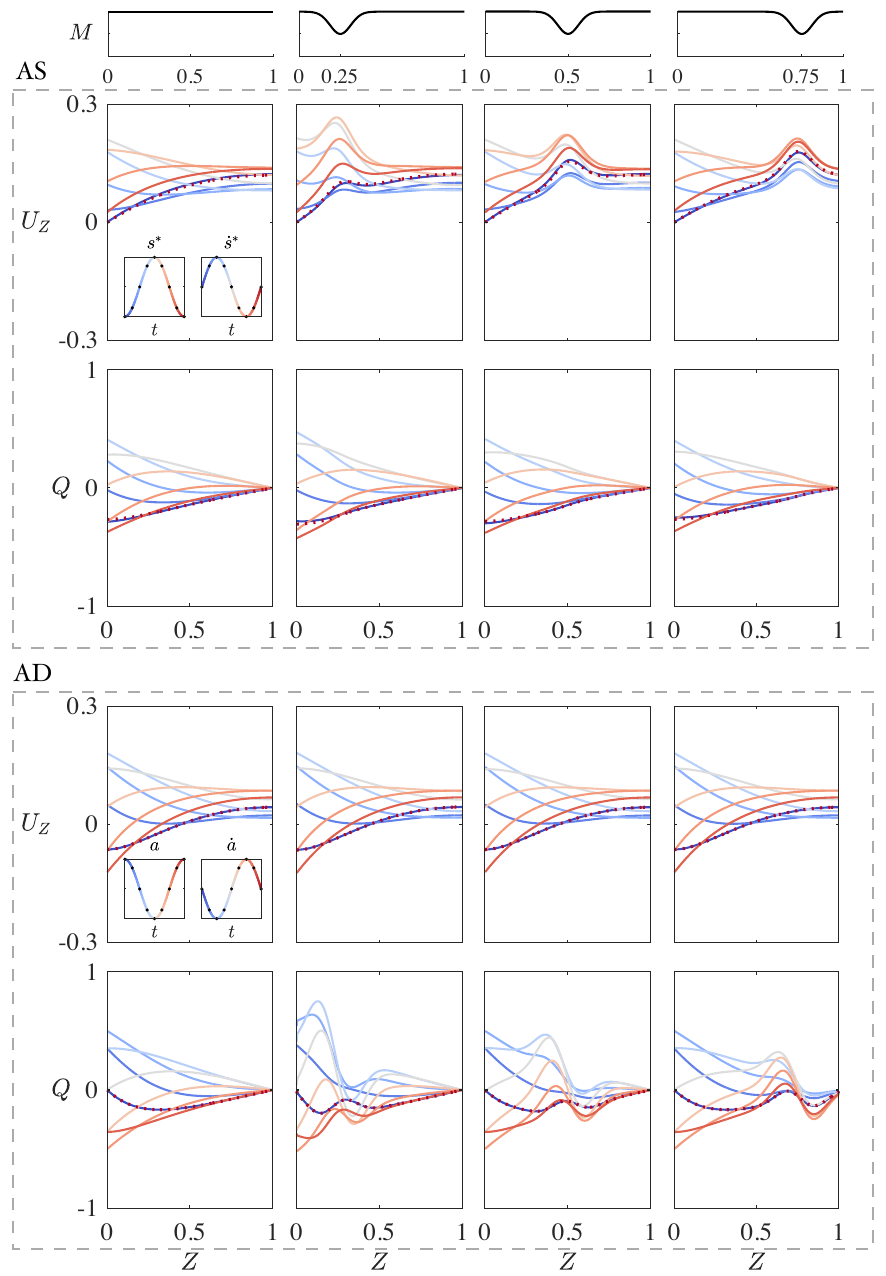}
    \caption{Strain (first and third rows) and flux (second and fourth rows) plotted for 9 equally spaced values of cycle $38\pi/\omega \leq t \leq 40\pi/\omega$, for uniform stiffness (first column) and locally decreased stiffness with $d=0.35$ (columns 2-4). The first two rows are in response to an applied stress ($A_l=0.2$, $\omega=10$) and the second two rows are in response to an applied displacement ($A_d=0.1$, $\omega=10$). The applied stress $s^*(t)$ and displacement $A(t)$, and their respective time derivatives $\Dot{s}^*(t)$, $\Dot{a}(t)$ are displayed as insets in the first column. We differentiate between the loading phase ($\Dot{s}^*>0$, $\Dot{a}<0$, dark blue to light blue) and unloading phase ($\Dot{s}^*<0$,$\Dot{a}>0$, light red to dark red). The end of the cycle is in dotted red.}
    \label{Fig: Strain and flux profiles with stiffness damage}
\end{figure}

In Figure \ref{Fig: Strain and flux profiles with stiffness damage} we plot the strain $U_Z(Z,t_i)$ and flux $Q(Z,t_i)$ profiles, at 9 time points, equally spaced over one period, under an applied stress (AS) with $(A_l,\omega)=(0.2,10)$ in the first two rows and under an applied displacement (AD) with $(A_d,\omega)=(0.1,10)$ in the last two rows. We differentiate between the loading phase, $\Dot{s}^*>0$ or $\Dot{a}<0$ (dark blue to light blue) and unloading phase, $\Dot{s}^*<0$ or $\Dot{a}>0$ (light red to dark red), with the end of the cycle in dotted red.  The first column is the homogeneous case, and columns 2-4 display profiles with imposed local decrease in stiffness where the location of the dip is increased from left to right (moving away from the moving boundary). Furthermore, $A_{l,d}$ and $\sqrt{T} = \sqrt{2\pi/\omega} $ are of the same order, and hence we are in a regime where we do not anticipate the linear model to provide an accurate approximation \cite{fiori:2023}.

The governing equation \eqref{Eq:1D Non-linear flow Summary} is a non-linear advection-diffusion equation (the advection term is zero for homogeneous stiffness) for porosity, or strain since $U_Z = \hat{\Phi}$; due to the constraint of uniaxial deformation and solid incompressibility, any increase in strain results in an increase in pore space. We refer to strain and porosity interchangeably in the discussion that follows. 

A key feature of the non-linear model is porosity-dependent permeability: with the Kozeny-Carman expression for permeability \eqref{Eq. perm law Summary} and neo-Hookean elasticity \eqref{Eq. stress law Summary}, the diffusion coefficient \eqref{Eq. Advection Diffusion Summary} takes the form
\begin{equation}
    D=\frac{k}{J}\frac{\partial s'}{\partial Z} = \frac{1}{2}\frac{(\hat{\Phi}/\Phi_0+1)^3}{(1+\hat{\Phi})^2}\biggl(M+\Lambda+\frac{M-\Lambda}{(1+\hat{\Phi})^2}\biggl),
\end{equation}
which is an increasing function of porosity, so long as $\Lambda < M$ (and $\hat{\Phi}>-\Phi_0$ i.e.\ $\Phi>0$ which is always true). The Kozeny-Carman term $k_{KC}$ increases with porosity faster than the stress term $J^{-1}\partial s'/\partial \hat{\Phi}$ decreases with porosity, so that diffusion as a whole is increased as porosity increases. Note that if we had constant $k=1$, the diffusion coefficient would be a decreasing function of porosity. 

The implications of the porosity-dependent permeability can be seen in Figure \ref{Fig: Strain and flux profiles with stiffness damage}. As the material is pulled (loading, blue lines, $\Dot{s}^*>0$, $\Dot{a}<0$), fluid starts to enter (positive flow), although this is not an instantaneous response, to satisfy the incompressibility condition (rows 2 and 4 of Fig.\ref{Fig: Strain and flux profiles with stiffness damage}), increasing porosity (rows 1 and 3) which increases permeability and thus diffusivity. The increased diffusivity means the stress imposed by AS or generated by AD at $Z=0$, and the associated strain, diffuses farther into the material over the cycle than it would if permeability did not depend on porosity. This also causes porosity to increase further, creating a positive feedback loop where increased porosity increases permeability which increases diffusivity which increases porosity and so on. As tensile stress decreases, or the material is pushed back (unloading, red lines, $\Dot{s}^*<0$, $\Dot{a}>0$), squeezing fluid out, diffusivity decreases (but not below $1$), slowing the response and diffusion of strain/porosity through the material. Naturally, strain still decays to some extent as $Z\to 1$, the extent of which depends on $\omega$, however it would decay much faster if permeability $k$ were constant. In addition, flux decays with $Z$ and is zero at the upper boundary $Z=1$ as required by Eq.\eqref{Eq. BC Z=1 Summary} (rows 2 and 4). 

Under AS, strain/porosity at the lower boundary $Z=0$ returns to zero at the end of the cycle, as depicted by the red dotted line in the first row of Fig. \ref{Fig: Strain and flux profiles with stiffness damage}. This is because $s'\approx M\hat{\Phi}$; in particular, this relation is exact for $s'=0$, i.e.\ when stress is zero, strain/porosity is zero. For $Z>0$ however, $\hat{\Phi}>0$, indicating stress is non-zero, even at the end of the cycle. This is due to the diffusive nature of the system: the stress acquired as $Z\to 1$ is not instantaneously lost. Additionally, whether the imposed stress at $Z=0$ eventually reaches the material points as $Z\to 1$ will depend on $\omega$. Since the solid phase is incompressible, $\hat{\Phi}>0$ indicates an increase in volume of the material; the fluid that entered during the transient period (initial cycles) is trapped for the duration of loading, and will only exit again once the loading stops and the material can fully relax. For our 1D system, an increase in volume is equivalent to a negative displacement at $Z=0$, so we can also write this as $U(0,t)<0$ for $Z>0$.

Under AD on the other hand (third row), we observe a negative strain near $Z=0$ at the end of each cycle (red dotted line), highlighting the compressive stress required to push the material back to its initial position (we note that displacement $U(0,t) = 0$ for $t=2n\pi/\omega$, integer $n$). The strain/porosity at the end of the cycle is also non-zero throughout the rest of the material, indicating that the material does not return to its initial configuration here either, despite returning to its initial volume. This is possible via a re-distribution of porosity throughout the material: we see that porosity is decreased near $Z=0$ and increased near $Z=1$ (such that total porosity $\int_0^1 \hat{\Phi}(Z,t=2n\pi/\omega)dZ = 0$ as required). Again the diffusive nature of the system is at play: the resulting stress/strain from the imposed displacement is not instantaneously transferred to the rest of the material, and how far it will penetrate the material over the cycle depends on $\omega$. The smaller $\omega$, the less compressive strain will be generated near $Z=0$ at the end of the cycle and the more homogeneous the strain will be in $Z$. For $\omega = 10$ here, an intermediate frequency, the response is not strongly localised to $Z=0$ but is not uniform in $Z$ either.

As expected under AS (first row of Fig.\ref{Fig: Strain and flux profiles with stiffness damage}), an overall decrease in stiffness results in greater strains, as the material is overall weaker, implying greater displacements. Strain is particularly increased at the point of maximum damage. For example, with damage imposed at $l=0.25$ (second column), we find a local increase in strain corresponding to the local decrease in stiffness, and exhibiting a maximum at $Z=0.25$.  The flux profiles (second row) are much less affected by the heterogeneity, showing only a small increase around $Z=l$. Though not shown in these plots, we also find that the transient phase is longer when stiffness is perturbed near the free boundary (small $l$). 

To understand how porosity can be increased whilst flux is negligibly affected, consider the expression \eqref{Eq: Q Vs} for flux, and let $\delta \Phi$ be a small increase in porosity. We can assume $\delta\Phi/J$ is small since $\delta\Phi\lessapprox0.1$ here and $J>1$. Let $J_\delta = J+\delta\Phi$ and $V_\delta^{f,s}$ be the new fluid and solid velocities. We find $Q_\delta = (V_\delta^f-V_\delta^s)(\Phi+\delta\Phi)/J_\delta =  (V_\delta^f-V_\delta^s)\Phi/J + O(\delta\Phi/J)$. Assuming $V^{f,s}_\delta \approx V^{f,s}$ then $Q_\delta \approx Q$. Since the velocity $V^{f,s}$ is unchanged, this equates to the ratio of porosity to volume change staying approximately constant for small $\delta \Phi/J$. This approximation breaks down as $\delta\Phi$ and $V^{f,s}_\delta$ grow, but for the small values dealt with here it is reasonable. Moreover, we note that this is Lagrangian flux, and the effect on true, Eulerian, flux $q$ will be greater since $q = JQ$, i.e.\ $q_\delta \approx q + Q\delta\Phi $.

Under AD on the other hand, we find that (for these parameters) strain, and by association displacement, is unaffected by the local decrease in stiffness (third row), whilst flux, and by association stress, is heavily impacted (fourth row). As evident in the last row, the flow is locally perturbed around the damage location, characterised by a local increase or decrease in inflow or outflow either side of $Z=l$ compared to the homogeneous case. For example for $l=0.25$ (2nd column), inflow is increased to the left of $l$ and decreased (or turned to outflow) to the right of $l$ for most of the cycle, however towards the end of the cycle outflow is decreased (or turned to inflow) to the left of $l$ and increased to the right of $l$. For $l=0.75$ on the other hand (4th column), inflow is always increased to the left of $l$ and outflow is always increased to the right of $l$. For all cases, the maximum magnitude of this \textit{increase} in inflow is located at $l-c$ and the maximum magnitude of increase in outflow is at $l+c$, where $c$ is the variance of the Gaussian defined in \eqref{Eq: Gaussian}. Note that flux is unchanged at $Z=0$, as required by \eqref{eq: Vs = -Q} and \eqref{Eq. BC Z=0 Summary} under AD.

\begin{figure}[h!]
    \centering
    \includegraphics[width=9cm]{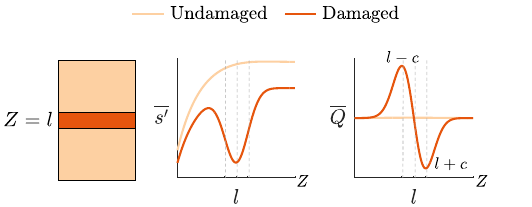}
    \caption{Time integrated stress and flux response for uniform stiffness (light
orange) and locally decreased stiffness (dark orange) against $Z$}
    \label{fig:schematic-stress-flux}
\end{figure}

To understand the effect on flux under AD, consider now the expression \eqref{Eq: Q dP} for $Q$. Flux is governed by stress gradients (recall $\partial s'/\partial Z=\partial P/\partial Z$) and permeability and porosity of the material ($k/(1+\hat{\Phi})$). Under an applied displacement, the position of the material at $Z=0$ is imposed, irrespective of its stiffness (and thus the flux at $Z=0$ too). If the material is locally weakened, a smaller stress is required to achieve the associated displacement throughout the material, and we see a local dip in the stress profile in the vicinity of the damage. This modifies the stress and thus pressure gradient, which modifies the flux profile accordingly, as illustrated in Figure \ref{fig:schematic-stress-flux} which plots the time integrated stress $\overline{s'}:=\int_0^{t_f}s'dt$ and time integrated flux $\overline{Q}:=\int_0^{t_f}Qdt$ for the homogeneous case in yellow and the damaged cases in orange against $Z$, where $t_f = 40\pi/\omega$. Since $\hat{\Phi} \approx s'/M$ and the ratio $s'/M$ remains approximately constant, strain/porosity is negligibly affected.

Additionally, we find that under both AS and AD the maximum strain/flux in the damaged area decreases with increasing $l$, however the magnitude of strain/flux in the damaged area is also more uniform in time for greater $l$ (rows 1 and 4). For example under AS, for $l = 0.25$, the magnitude of strain at $l$ varies between $0.08$ and $0.27$, whereas for $l=0.75$, it only varies between $0.13$ and $0.21$. Similarly under AD, for $l = 0.25$, the magnitude of flux at $l-c$ varies between $-0.35$ and $0.73$, whereas for $l=0.75$, it only varies between $-0.25$ and $0.06$. This is once again due to the diffusive nature of the system which localises the stress-strain response to $Z=0$ where the boundary condition is applied, so that stress and strain vary more near $Z=0$ and less as $Z\to1$.

\subsubsection{Defining a metric}

The above results highlight that heterogeneous stiffness impacts the strain and flux of the material under AS/AD both temporally and spatially. To characterise the system response in a more concise form, in this section we define and analyse several metrics that capture the net response. To this end, we first define the following time integrated quantities:
\begin{equation}\label{eq. time integrated}
    \overline{|U_Z|} = \int_0^{t_f} |U_Z|dt, \quad \overline{|Q|} = \int_0^{t_f} |Q| dt,
\end{equation}
where $t_f=40\pi/\omega$. Note this definition includes the transient period. By integrating the modulus, these metrics do not differentiate between compressive and tensile strain, or between inflow and outflow. Under AS, the absolute value has no effect on strain since $U_Z\geq 0$. Under AD on the other hand the compressive strain now has an \textit{additive} effect with tensile strain. In both cases, outflow and inflow are also additive. These metrics are therefore cumulative, and the more cycles are applied, the more the effect of damage on strain and flux will cumulate, increasing the difference with the undamaged case. We refer to $\overline{|U_Z|}$ as “cumulative strain” and $\overline{|Q|}$ as “cumulative flux”. 

\begin{figure}[h!]
\centering
\includegraphics[width=9cm]{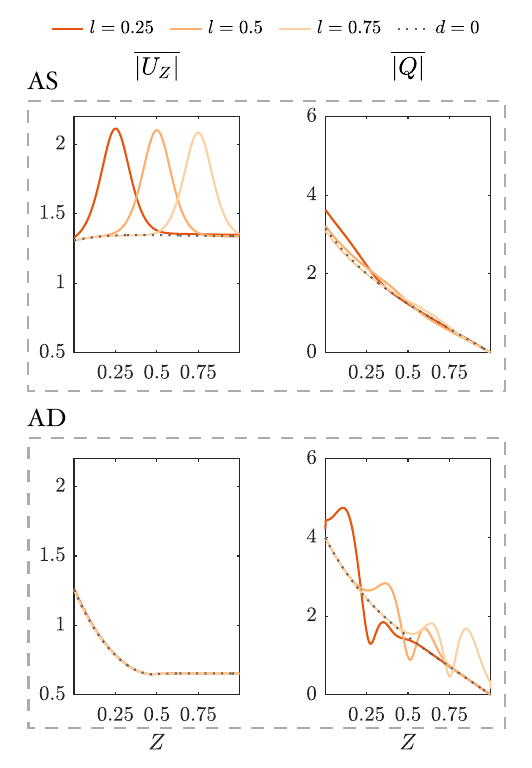}
    \caption{Cumulative strain (left) and flux (right) against $Z$ under AS ($A_l=0.2$, $\omega=10$, top row) and AD ($A_d=0.1$, $\omega=10$, bottom row), for homogeneous stiffness (dotted grey line) and for damaged stiffness with $d = 0.35$ and varying location $l = 0.25,0.5,0.75$.}
    \label{Fig: Time integrated strain and flux profiles with stiffness damage}
\end{figure}

The cumulative strain and flux are plotted in Figure \ref{Fig: Time integrated strain and flux profiles with stiffness damage} for $d=0.35$ and three locations $l$ (increasing $l$ from dark to light orange) as well as the uniform case in dotted grey. The top left panel highlights the increase in strain about $Z=l$ under AS, and the bottom right panel highlights the disturbed flux about $Z=l$ under AD. The figure additionally highlights the negligible effect heterogeneous stiffness has on strain under AD (bottom left) and the small effect on flux under AS (top right).

Clearly, the spatial features present in the strain and flux fields in Figure \ref{Fig: Strain and flux profiles with stiffness damage} are not necessarily preserved in the cumulative flux and strain profiles. For example, for $l=0.25$, $\overline{|Q|}$ is decreased to the right of $Z=l$. For $l=0.5$ and $l=0.75$, the cumulative flux to the right of $l$ is first decreased and then increased. Additionally, the cumulative strain under AS exhibits the same profile and magnitude for each value of $l$.

A natural extension to this spatial metric is to integrate over $Z$ as well:
\begin{equation}
    <\overline{|U_Z|}> = \int_0^1\int_0^{t_f} |U_Z| dtdZ, \quad <\overline{|Q|}> = \int_0^1\int_0^{t_f} |Q| dtdZ.
\end{equation}
where the time integral captures the cumulative effect of damage on strain or flux magnitude as before, and the spatial integral averages the cumulative strain/flux over space. We refer to these quantities as “net strain” and “net flux” respectively. In Figure \ref{Fig: Time integrated strain and flux profiles with stiffness damage} this simply corresponds to the area under the $\overline{|U_Z|}$ and $\overline{|Q|}$ curves. For this particular case, we find that although the maximum strain/flux about $Z=l$ decreases with $l$, the net strain  is approximately equal for all $l$ under AS and the net flux slightly increases with $l$ under AD. As these metrics output a single number for any given form of heterogeneity and loading conditions, they enable to more readily explore the system response to variation in parameters, which we do in the following subsection. 

\subsubsection{Parameter variation}

In order to more thoroughly investigate the effect of damage magnitude $d$, damage location $l$, and loading frequency $\omega$, we continuously vary $d$ and $\omega$ for three values of $l$, computing the net strain and flux for each parameter set. We then compute the difference between the damaged and undamaged net strain/flux via 
\begin{equation} 
\begin{split}
    \Delta U = \;<\overline{|U_{Z,d}|}> &- <\overline{|U_{Z,u}|}>\;, \\
    \Delta Q = \;<\overline{|Q_d|}> &- <\overline{|Q_u|}>,
\end{split}
\end{equation}
where $\cdot_d$ and $\cdot_u$ denote “damaged” and “undamaged” respectively. If $\Delta U,\Delta Q>0$ then damage increases net strain/flux, and conversely if $\Delta U, \Delta Q<0$ then damage decreases net strain/flux. An increase in $|\Delta U|$ or $|\Delta Q|$ means an increase in the effect of damage.

\begin{figure}[h!]
    \centering
\includegraphics[width=18cm]{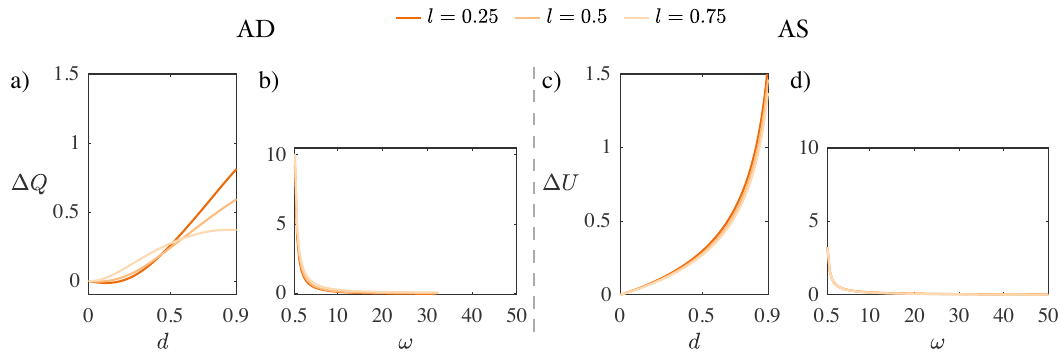}
    \caption{Difference between damaged and undamaged net flux under AD ($A_d=0.1$) against: a) stiffness damage magnitude $d$ (fixed $\omega = 10$; b) loading frequency $\omega$ (fixed $d=0.35$), for three values of $l$. Difference between damaged and undamaged net strain under AS ($A_l=0.2$) against: a) stiffness damage magnitude $d$ (fixed $\omega = 10$; b) loading frequency $\omega$ (fixed $d=0.35$), for three values of $l$. }
    \label{Fig: Effect-params-stiff}
\end{figure}

Unsurprisingly, increasing damage magnitude $d$ increases the effect of damage. For example the difference in net flux $\Delta Q$ under AD increases with $d$, as illustrated in Figure \ref{Fig: Effect-params-stiff} a). Here, the dependence of net flux on damage location $l$ is not the same for all $d$: for $d\leq0.5$ net flux increases with $l$, in line with our observations from the previous section, whilst for $d\geq0.5$ the relationship reverses such that net flux decreases with $l$. This can be understood by noting that as $d$ increases, the decrease with $l$ in maximum flux at $l\pm c$ steepens, such that beyond $d=0.5$ the net flux also decreases with $l$. In most cases however $\Delta U$ and $\Delta Q$ decrease with $l$, as illustrated in \ref{Fig: Effect-params-stiff} c) for $\Delta U$ under AS or in Appendix \ref{App: Param var stiff}.

The effect of stiffness damage on net flux under AD increases exponentially with decreasing frequency, as illustrated in Figure \ref{Fig: Effect-params-stiff} b), whilst net strain remains small for all $\omega$ (see Appendix \ref{App: Param var stiff - AD}). The simulations were run for $0.5\leq\omega\leq50$ however the curves stop at $\omega = 32.5$, beyond which porosity locally vanishes and the code breaks. This is a known characteristic of a poroelastic material subject to cyclic applied displacement; we refer the reader to \cite{hewitt:2016} and \cite{fiori:2023,fiori:2024} for further details. Also noteworthy is the fact that our simulation time is based on a fixed number of loading cycles, therefore a smaller $\omega$ increases the total loading time. The same trend of increased effect at smaller frequency is seen for $\Delta U$ under AS in \ref{Fig: Effect-params-stiff} d), as well as for $\Delta Q$ (see Appendix \ref{App: Param var stiff - AS}), although smaller in magnitude (note vertical scale). 

This increase in net strain/flux for decreasing $\omega$ is due to the increased loading period and the homogenisation of the response in $Z$. At large frequency, the strain and flux remain localised to the moving boundary, but for smaller frequency, as the loading increases, the strain tends to be more uniform in $Z$, homogenising the response, and the flux tends to linear in $Z$. The response of the material penetrates further into the material away from the boundary, leading to greater cumulative strain and flux, which is further amplified by the increased loading period. However, $\Delta U$ under AD shows a non-monotonic dependence on $\omega$, though the magnitude is much smaller ($\Delta U\sim10^{-2}$, see Appendix \ref{App: Param var stiff - AD}).  A summary of the response of $\Delta U$ and $\Delta Q$ under different parameter changes and loading regimes is presented in Appendix \ref{App: Param var stiff - table}.

Another parameter that is natural to vary is loading amplitude, but as previously discussed loading amplitude only has a quantitative and intuitive impact, so that increasing loading amplitude $A_l$ or $A_d$ will only amplify the effects of heterogeneity, and vice versa.

\subsection{Heterogeneous permeability}\label{Sec: Het perm}

\begin{figure}[h!]
    \centering
\includegraphics[width=17cm]{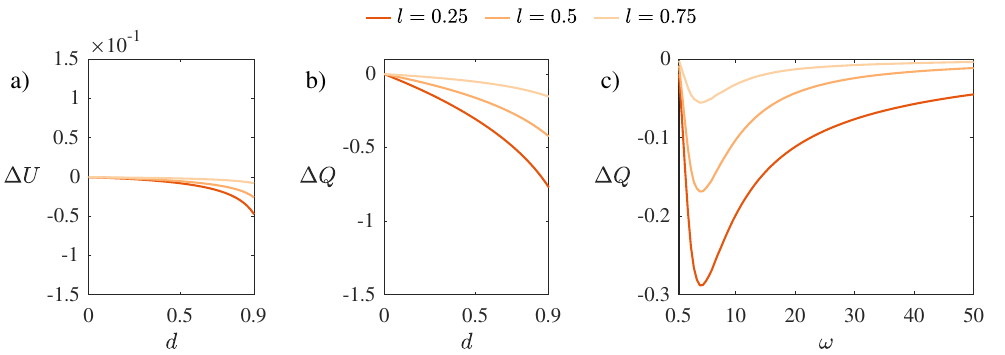}
    \caption{a) Difference between damaged and undamaged net strain under AS ($A_l=0.2$) against permeability damage magnitude $d$ (fixed $\omega = 10$).  b-c) Difference between damaged and undamaged net flux under AS ($A_l=0.2$) against b) permeability damage magnitude $d$ (fixed $\omega = 10$) and c) loading frequency $\omega$ (fixed $d=0.35$), for three values of $l$.}
    \label{Fig: Effect-params-perm}
\end{figure}

So far we have only considered heterogeneous stiffness, however heterogeneity may be introduced in other material properties, such as permeability $k = k_{KC}f(Z)$. Figure \ref{Fig: Effect-params-perm} displays the effect of permeability damage magnitude $d$ on the net strain and net flux and the effect of loading frequency on net flux under AS (see Appendix \ref{App: perm - full solution} for the full solution -- the spatial profiles plotted at time points over a loading cycle -- with $d=0.8$, $\omega=10$). Decreasing permeability naturally reduces the flow (c.f.\ equation \eqref{Eq: Q dP}) as shown by $\Delta Q<0$. It also decreases diffusion, so that strain is decreased under AS ($\Delta U<0$) and increases \textit{compressive} strain under AD ($\Delta U>0$, not shown). Again, $|\Delta U|$ and $|\Delta Q|$ increase with $d$ and decrease with $l$. The dependence on $\omega$ however is different to stiffness damage, displaying a non-monotonic profile as illustrated in the right panel of Fig. \ref{Fig: Effect-params-perm}. This is a result of $Q$ both increasing in magnitude and becoming more localised to $Z=0$ as $\omega$ increases, maximising the effect of permeability damage at $\omega = \omega^*$. In this case, $\omega^*\approx1.5$ and the value of $\omega^*$ varies for AS, AD, $\Delta U$ and $\Delta Q$. A summary of the response of $\Delta U$ and $\Delta Q$ under different parameter changes and loading regimes is presented in Appendix \ref{App: perm - table}.

Whilst the order of magnitude of $\Delta Q$ is comparable between permeability and stiffness damage when varying $d$, net flux is considerably more sensitive to loading frequency with stiffness damage (subject to their different loading strategies). Additionally, the effect of permeability damage on net strain is one to two orders of magnitude smaller than with stiffness damage. When looking at the full solution over a steady cycle however (Appendix \ref{App: perm}), the strain's qualitative profile is considerably more affected than the flux's, highlighting the care that must be taken when interpreting results from a metric such as net strain/flux, particularly if trying to infer information about the full behaviour of the material. Note that to accentuate the qualitative effect of permeability damage, Appendix \ref{App: perm} plots the full solution for the larger value $d=0.8$ than was used for stiffness damage.

\section{Discussion and Biological Relevance}

In this paper, we have investigated the effect of continuous material heterogeneity on the response of a poroelastic material to a uniaxial applied stress or applied displacement. While incorporating heterogeneity and cyclic loading creates a high-dimensional parameter space, we generally found that decreased stiffness increases strain under AS and increases flux under AD, whereas decreased permeability decreases flux under AS and increases absolute strain under AD, though the effect of permeability damage is one to two orders of magnitude smaller than stiffness damage. We also examined how the net magnitude of the strain and flux response varies with damage magnitude, location and frequency of AS/AD.

As our analysis was motivated by damage in tendon, it is worth interpreting our results in this context. In soft hydrated tissues such as tendon or cartilage, cells respond to their mechanical environment. Assuming for simplicity a symmetry between tension and compression (or flux direction),  the cumulative (time integrated) strain or flux provide reasonable measures for the local mechanical environment experienced by a cell. Consider a tendon subject to cyclic applied loads, with damage creating a local decrease in stiffness, and suppose for instance that the tendon cells' response to their mechanical environment correlates with cumulative strain \cite{nakamichi:2024,humphrey:2001}. In this scenario, based on our analysis, the greater cumulative strain (compared to an undamaged tendon) would indicate a modified protein production (mechanotransduction) which in turn would modify the structure and thus material properties of the tendon, feeding back into how the tendon responds to the mechanical stress. 

While our model of course presents a highly idealized system that does not include much of the biological or structural complexity of a real tendon, the response features we have uncovered could in principle constitute a basic diagnostic tool indicating how severe damage is and its location, given baselines values of strain and flux for healthy tissue. Our results suggest that the flux gives a better indication on location of damage, and that slower loading will result in a greater difference in net strain/flux between the healthy and unhealthy tendon. In the context of tendon mechanics, the regimes of fast, moderate and slow loading correspond to different activities: for the Achilles tendon for example, fast loading would correspond to running, moderate loading to walking or slow jogging, and slow loading to exercises such as resistance training. Given $T_{pe} \approx 14\unit{s}$ for our choice of parameters, we identify slow loading as $\omega \leq 2$ \cite{beyer:2015}, moderate loading as $2\leq \omega \leq 20$, and fast loading as $\omega \geq 20$ \cite{rowlands:2007}, which are all regimes we have considered in our study. Note that amplitude of loading also varies with these different exercises; although we did not vary amplitude in this study, its qualitative effect in our regime of interest is negligible for the homogeneous case \cite{fiori:2023} and we can expect a similar result for heterogeneous stiffness/permeability.  Additionally, since the maximum strain magnitude decreases with damage location $l$, a measurement of maximum strain could give an indication of the damage location.

In our analysis we used the net strain and flux to characterise how the response varies with damage magnitude, location and frequency. This was primarily motivated by the form of the solution which, except for strain under AS, oscillates between positive and negative values over a cycle, and thus cancels out (to some extent) when integrating over time. The idea was to capture the cumulative effect of damage on the magnitude of the response. In practice, the choice of metric should be informed by the context of the problem. For example, the mechano-sensitive response of tendon cells may in fact correlate more closely with maximum strain or flux, or stress gradients, or net pressure. The sign of these measures might also be important, responding differently in relation to tensile and compressive strain, or to inflow versus outflow. In problems concerned with solute transport, the difference between positive and negative flux is clearly key. In practice, the choice of measure is also limited by the tools available. For instance in a medical exam measuring the strain at each $Z$ might not be possible, however measuring macroscopic strain may be more accessible. This separates research aimed at improving our understanding of the development of diseased or damaged tissue, to inform prevention and healing, and research aimed at improving diagnosis.

It is also worth noting that some care should be taken when comparing results to experimental data, as we have shown that applying a stress or applying a displacement generate very different responses, particularly when heterogeneity is imposed. For example, \textit{in vivo} it might be more natural to apply a stress, or even a combination of stress and displacement due to the anatomical constraints, whereas \textit{ex vivo} or \textit{in vitro} experiments are often performed by means of an applied displacement as this is easier to manipulate.  It should therefore not come as a surprise if \textit{in vivo} and \textit{ex vivo}/\textit{in vitro} experiments yield different results, and comparison to theory should be conducted accordingly.

    \section{Conclusion}

We have presented a fluid-saturated poroelastic model in a Lagrangian framework for a material undergoing uniaxial cyclic tensile stress or displacement, where the solid obeys neo-Hookean elasticity and permeability is porosity-dependent according to Kozeny-Carman's law. The novel contribution from our study is the elucidation of the effect of continuous heterogeneous material properties, referred to as “damage” and imposed as a local decrease in the stiffness or permeability by a magnitude $d$ at a location $l$, on the strain and flux response to an applied stress or displacement. Although damage was found to affect the strain and flux response both spatially and temporally, we were able to reduce the response to a pair of scalar metrics to explore the effect of loading frequency and damage parameters $d$ and $l$. While the metrics omitted the spatial and temporal detail of the response they gave a general sense of the system's sensitivity to these parameters and can serve as a useful guide for more specific studies. A further strength of our modelling approach is, given agreement between our predictions for net quantities and experimental data, the model also determines the underlying spatial and temporal features of the fields of interest.

The results from our analysis provide an insight into how material heterogeneity modifies the response of soft porous materials to cyclic loads or displacements. While the quantitative details will depend on modelling choices -- in particular the form of heterogeneity and the  metrics analysed -- the qualitative principles we have uncovered are generally applicable to a wide range of physical systems, from rock mechanics to hydrated tissues. We were particularly motivated by the development of disease such as tendinitis and tendinopathy, in which the material properties of the tendon such as stiffness and permeability are altered. We chose to directly damage permeability, however we could also have indirectly manipulated it by damaging the initial porosity $\Phi_0$ instead -- to fully understand the relationship between modifying porosity and the associated permeability of a poroelastic material would require a multi-scale model. The literature also suggests tendons are highly anisotropic and that fluid exchange is possible in the transverse directions, so that future work could be directed towards constructing a reduced three-dimensional model which takes into account these effects. Moreover, the neo-Hookean model and the Kozeny-Carman expression for permeability were chosen for their simplicity and wide use, and substitution with more relevant constitutive laws for the material in question in this model should be straightforward, provided thermodynamic constraints are respected. For instance, a tendon specific constitutive equation for the solid skeleton could be employed instead. The question of which mechanical cues are relevant to mechanotransduction remains open and is an active area of research, and will help further inform future work on this macroscopic model of the tendon. We highlight that the simplicity of this model provides us with a clear and intuitive understanding of the behaviour of heterogeneous poroelastic materials, which can be used to inform and build more complex, higher-dimensional models and incorporate additional biology.

\section*{Author contributions}
Designed research: all authors;
Conducted research: Z.C.G.;
Wrote paper: all authors

\section*{Conflicts of interest}
There are no conflicts to declare.

\section*{Data availability}
The code required to reproduce model output will be made available in a public repository upon acceptance.


\section*{Acknowledgements}
The authors thank Chris MacMinn for helpful discussions. 

\section*{Funding}
Z.C.G. gratefully acknowledges funding from the UKRI and the Paul Shreder scholarship. For the purpose of open access, the author has applied a CC BY public copyright licence to any author accepted manuscript arising from this submission.


\appendix

\section{Surface transport equations}\label{app: Surf trans}

We derive two useful results which relate the flow of Eulerian quantities through a material surface $da$ to the flow of Lagrangian quantities through a material surface $dA$. Nanson's formula states that a material surface $dA$ with unit normal $\mathbf{N}$ transforms to the material surface $da$ with unit normal $\mathbf{n}$ as
\begin{equation}\label{Eq. Nanson}
    \textbf{n}da = J\textbf{F}^{-T}\textbf{N}dA \quad\text{or}\quad n_ida = JF^{-1}_{ji}N_jdA.
\end{equation}
Consider an arbitrary vector $\mathbf{M}$ in the reference configuration which transforms to the vector $\mathbf{m}$ attached to the current configuration, such that the flow of $\mathbf{M}$ through $dA$ is equivalent to the flow of $\mathbf{m}$ through $da$:
\begin{equation}\label{Eq. m.dot.n = M.dot.N}
    \mathbf{m}\cdot\mathbf{n}\;da = \mathbf{M}\cdot\mathbf{N}\;dA.
\end{equation}
Using Nanson's formula this can be rearranged as
\begin{equation}\label{Eq. Nanson for M}
    \mathbf{M} = J\mathbf{F}^{-1}\mathbf{m} \quad\text{or}\quad M_I=J\frac{\partial X_I}{\partial x_j}m_j.
\end{equation}
Moreover, integrating \eqref{Eq. m.dot.n = M.dot.N} and applying the divergence theorem for arbitrary volume $dv$ and $dV_0$:
\begin{equation}\label{Eq. Nanson for dM}   \nabla_X\cdot\mathbf{M}\;dV_0=\nabla_x\cdot\mathbf{m}\;dv.
\end{equation}
Equation \eqref{Eq. Nanson for M} is useful when converting relevant Eulerian quantities to Lagrangian ones, and equation \eqref{Eq. Nanson for dM} will be useful when converting from Eulerian gradient to Lagrangian gradient, provided Eq.\eqref{Eq. m.dot.n = M.dot.N} holds.

\section{Numerical Method}\label{App: method}

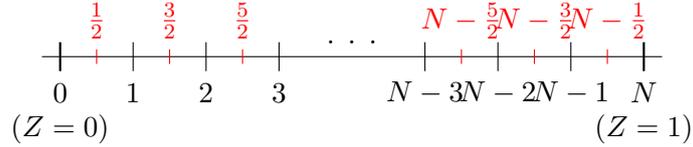
\begin{figure}[h]
    \centering
    \begin{center}
\begin{tikzpicture}[scale = 0.48]

\draw[black] (-8.5,0) -- (8.5,0);
\draw[black,thick] (-8,-0.4) -- (-8,0.4);
\node at (-8,-1) {$0$};
\draw[black,thick] (8,-0.4) -- (8,0.4);
\node at (8,-1) {$N$};
\draw[black] (-6,-0.4)--(-6,0.4);
\node at (-6,-1) {$1$};
\draw[black] (-4,-0.4)--(-4,0.4);
\node at (-4,-1) {$2$};
\draw[black] (-2,-0.4)--(-2,0.4);
\node at (-2,-1) {$3$};
\draw[black] (6,-0.4)--(6,0.4);
\node at (6,-1) {$N-1$};
\draw[black] (4,-0.4)--(4,0.4);
\node at (4,-1) {$N-2$};
\draw[black] (2,-0.4)--(2,0.4);
\node at (2,-1) {$N-3$};
\draw[red] (-7,-0.2)--(-7,0.2);
\node[red] at (-7,1) {$\frac{1}{2}$};
\draw[red] (-5,-0.2)--(-5,0.2);
\node[red] at (-5,1) {$\frac{3}{2}$};
\draw[red] (-3,-0.2)--(-3,0.2);
\node[red] at (-3,1) {$\frac{5}{2}$};
\draw[red] (7,-0.2)--(7,0.2);
\node[red] at (7,1) {$N-\frac{1}{2}$};
\draw[red] (5,-0.2)--(5,0.2);
\node[red] at (5,1) {$N-\frac{3}{2}$};
\draw[red] (3,-0.2)--(3,0.2);
\node[red] at (3,1) {$N-\frac{5}{2}$};
\node at (0,0.4) {. . .};
\node at (-8,-2) {$(Z=0)$};
\node at (8,-2) {$(Z=1)$};
\end{tikzpicture}
\end{center}
    \caption{Finite Volume Mesh for a one-dimensional poroelastic material, with stress or displacement applied at $Z=0$ and no flux at $Z=1$. Unknown values $\bar \Phi_i$ are solved at each of the red nodes.}
    \label{1D_mesh}
\end{figure}

This numerical method is implemented to solve the advection-diffusion equation for porosity. Subsequent fields, such as pressure, displacement, stress and strain, are calculated after. Consider the finite volume mesh where the spatial domain $[0,1]$ is divided into $N$ grid cells, where the cell walls are located at the black mesh points $0\leq i \leq N$, and the cell centres are located at the red mesh points $\frac{1}{2} \leq i+\frac{1}{2} \leq N-\frac{1}{2}$, as depicted in Figure \ref{1D_mesh}. Let $Z_i = i\Delta Z$ where $\Delta Z =1/N$ so that $Z_0=0$ and $Z_N=1$, and denote the unknowns $\hat{\Phi}(Z_i,t)=\hat{\Phi}_i(t)$. We solve for $\hat{\Phi}_{i+1/2}(t)$ at the cell centres, and boundary conditions are applied at $i=0$ and $i=N$. 

Rewriting \eqref{Eq:1D Non-linear flow Summary} in the following form,
\begin{equation}\label{Eq. PDE numerical}
    \frac{\partial \hat{\Phi}}{\partial t} = \frac{\partial}{\partial Z}\biggl[G(\hat{\Phi})\frac{\partial s'}{\partial Z}\biggl],
\end{equation}
where $G(\hat{\Phi}) = k(\hat{\Phi})/(1+\hat{\Phi})$ we integrate the above equation over cell $i$ and divide by $\Delta Z$:
\begin{equation} 
    \frac{1}{\Delta Z}\int_{Z_{i}}^{Z_{i+1}} \frac{\partial \hat{\Phi}(Z,t)}{\partial t} dZ - \frac{1}{\Delta Z}\left[G(\hat{\Phi})\frac{\partial s'(Z,t)}{\partial Z}\right]^{Z_{i+1}}_{Z_{i}} = 0.
\end{equation}
Let the local cell average  $\bar \Phi_{i+\frac{1}{2}}(t)$ be given by 
\begin{equation}
    \bar \Phi_{i+\frac{1}{2}}(t) = \frac{1}{\Delta Z}\int_{Z_i}^{Z_{i+1}}\hat{\Phi}(Z,t)dZ, 
\end{equation}
and the corresponding stress $\bar s_{i+1/2}=s'(\bar \Phi_{i+1/2})$, calculated using equation \eqref{Eq. stress law} with $J_i = 1+\bar{\Phi}_i$. Let the numerical finite volume flux at the cell walls $F_i(t)$ be given by
\begin{multline}
    F_i(t) \equiv F(Z_i,t) = -G(\hat{\Phi}_i)\frac{\partial s'_i}{\partial Z} \\ \approx -G\left(\frac{\bar \Phi_{i+1/2}+\bar \Phi_{i-1/2}}{2}\right)\frac{\bar s_{i+1/2} - \bar s_{i-1/2}}{\Delta Z}
\end{multline}
where we have used a central average to approximate $D(\Phi_i)$ and a centered finite difference to approximate the derivatives for inner mesh points $1\leq i\leq N-1$. The PDE \eqref{Eq. PDE numerical} is now a system of ODEs for each timestep,
\begin{equation}
    \frac{d \bar \Phi_{i+1/2}}{dt} = -\frac{F_{i+1}-F_{i}}{\Delta Z}
\end{equation}
which we can solve using \texttt{ode15s} in MATLAB with the appropriate boundary and initial conditions. 
The above formulation can also be used for the form \eqref{Eq. non-linear flow equation} of the non-linear flow equation, giving 
\begin{equation}\label{Eq. FV fluxes Q}
    F_i(t) = - Q(Z_i,t) = V^s(Z_i,t)
\end{equation}
which is useful when implementing boundary conditions in the following section.

\subsection{Implementation of boundary and initial conditions}

\subsubsection{Zero flux and solid displacement}

The zero flux boundary condition at $i=N$ \eqref{Eq. BC Z=1} can be rewritten as $Q(Z_N,t) = 0$, so that from \eqref{Eq. FV fluxes Q}
\begin{equation}
    F_N(t) = 0.
\end{equation}

\subsubsection{Applied stress}

For an applied stress $s^*(t)$ at $Z=0$, we approximate the derivative $\partial s'/\partial Z$ using a forward finite difference such that
\begin{equation}
\frac{\partial s'_0}{\partial Z} \approx \frac{\bar s_{1/2} - s^* }{\Delta Z / 2}
\end{equation}
where the corresponding value $\Phi^* \coloneqq \hat{\Phi}(0,t)$ associated with $s^*$ is calculated by inverting \eqref{Eq. stress law} and choosing the positive root (this is the only physical choice for $\nu < 1/2$). The finite volume flux at $Z=0$ is then:
\begin{equation}F_0(t) \approx - G(\Phi^*)\frac{\bar s_{1/2} - s^* }{\Delta Z / 2}.
\end{equation}

\subsubsection{Applied displacement}

For an applied displacement at $Z=0$, we make use of the form \eqref{Eq. FV fluxes Q} so that
\begin{equation}
F_0(t) = \dot a.
\end{equation}

\subsubsection{Initial condition}

The ODE time solver is initialised with $\bar{\Phi}_i(0) = 0$.

\subsection{Other quantities}

Displacement is calculated from porosity by integrating Eq. \eqref{Eq. Uz=Phi)}:
\begin{equation}
    U(Z,t) = \int_1^Z \hat{\Phi}(Z,t) dZ.
\end{equation}
Pressure is calculated by integrating Eq. \eqref{Eq.MechEq}:
\begin{equation}
    P(Z,t) = s'(Z,t)-s^*(t)
\end{equation}
or in the case of applied displacement:
\begin{equation}
P(Z,t) \approx s'(Z,t) - s'(Z_{\frac{1}{2}},t),
\end{equation}
which is approximated since $s'(Z_0,t)$ cannot be computed, and stress is calculated from porosity using \eqref{Eq. stress law Summary}. Strain is simply given by
\begin{equation}
U_Z = \hat{\Phi}
\end{equation}
and the relative flow 
\begin{equation}
Q = -\frac{k(\hat{\Phi})}{J}\frac{\partial s'}{\partial Z} = -V^s
\end{equation}
which quantifies how much fluid is being transported through the material at a particular point. 

\subsection{Numerical convergence}

To determine an appropriate step size, we solve the system under an applied stress and varying the grid step-size $dZ = 1/N$ from $25$ to $1000$. We calculate the absolute relative error as
\begin{equation}
\epsilon_{rel} = \left|\frac{U_{current}(0,t_{max})-U_{old}(0,t_{max})}{U_{current}(0,t_{max})}\right|,
\end{equation}
evaluating displacement at the applied stress boundary ($Z=1/2N$) as this involves integrating $\Phi$ over the whole domain. For applied displacement we perform the same error calculation but evaluating stress at $Z=1/2N$.
\begin{figure}[h]
    \centering
    \includegraphics[width=8.2cm]{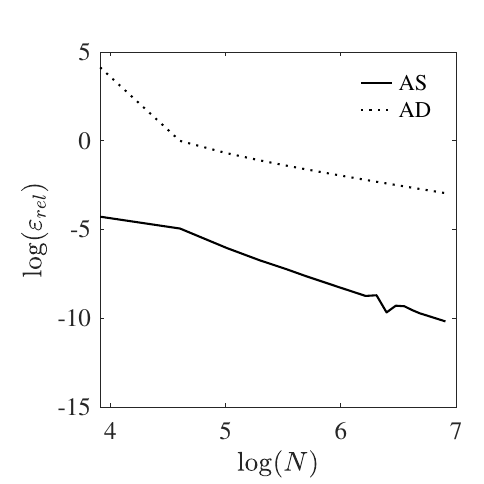}
    \caption{Relative error calculated for displacement decreasing the step-size (increasing N) for applied stress ($A_l = 0.2$, $\omega = 10$) and applied displacement ($A_l = 0.2$, $\omega = 10$).}
    \label{Fig. error stress}
\end{figure}

The relative error decreases with $N$ and we compute our results with $N=400$ ($\log(N)\approx 6$) so as not to compromise on computation speed. We note that the method to enforce the stress boundary condition converges much faster than the displacement boundary condition.

\section{Increased stiffness}\label{App. increased stiffness}

\begin{figure}[h!]
\centering
\includegraphics[width=9cm]{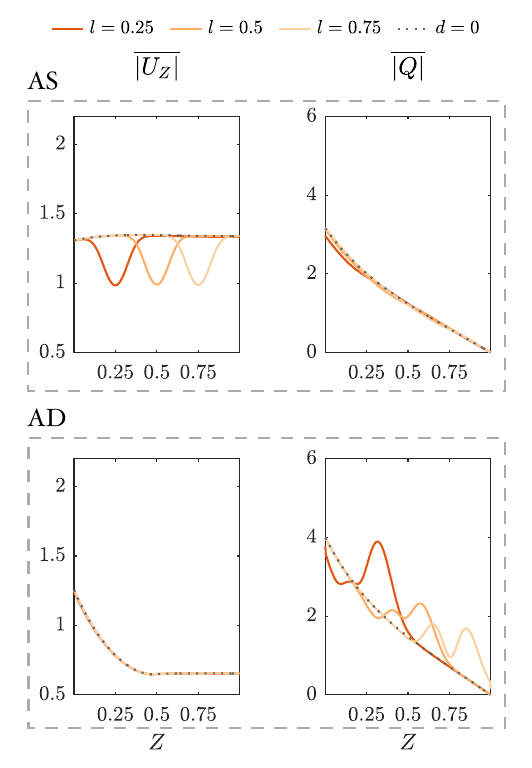}
    \caption{Time integrated absolute strain and flux profiles against $Z$ under AS ($A_l=0.2$, $\omega=10$, top row) and AD ($A_d=0.1$, $\omega=10$, bottom row), for homogeneous stiffness (dotted grey line) and for local increase in stiffness with $d = 0.35$ and varying location $l = 0.25,0.5,0.75$.}
    \label{Fig: Time integrated strain and flux profiles with stiffness increase}
\end{figure}

Consider a local \textit{increase} in stiffness instead of a decrease (to do so we simply change the sign of the Gaussian in Eq.\eqref{Eq: Gaussian}). This means that the material is locally stiffer, thus requiring more stress to be pulled and compressed to the same displacement. Under AS we can expect smaller strain values at the point of damage due to the stiffer region. Under AD, we expect a flux disturbance which mirrors the one seen for the local decrease in stiffness.

In Figure \ref{Fig: Time integrated strain and flux profiles with stiffness increase} we plot the time integrated absolute strain and flux under AS (first row) and under AD (second row) for a locally \textit{increased} stiffness of magnitude $d=0.35$ and locations $l = 0.25,0.5,0.75$ (to do so we simply change the sign of the Gaussian in Eq.\eqref{Eq: Gaussian}). This mirrors the effect seen for a local decrease in stiffness, whereby strain is locally decreased under AS, and flux inversely perturbed under AD about $Z=l$. We note the magnitude of the response is smaller compared to decreased stiffness. Computing the areas under these curves, i.e.\ the net absolute strain and flux, we find the net strain to be approximately equal for the three locations under AS, though this time it is slightly increased for increasing $l$ (whilst for decreased stiffness it was slightly increased for decreasing $l$). Under AD, the effect of location is not linear as before: $<\overline{|Q|}>$ is greatest for $l=0.75$ but smallest for $l=0.5$.

\section{Additional parameter variation for stiffness damage}\label{App: Param var stiff}

\subsection{Applied Displacement}\label{App: Param var stiff - AD}

\begin{figure}[h!]
    \centering
\includegraphics[width=9cm]{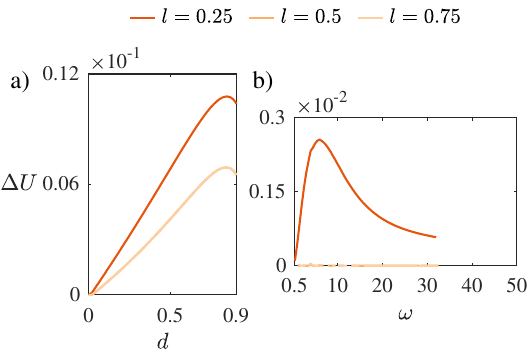}
    \caption{ Difference between damaged and undamaged net strain under AD ($A_d=0.1$) against a) $d$  and b) $\omega$ for three locations $l$.}
    \label{Fig: Effect params AD Appendix}
\end{figure}

Figure \ref{Fig: Effect params AD Appendix} displays the effect of damage magnitude $d$ and loading frequency $\omega$ on the net strain under AD. Clearly, the effect of damage on net strain under AD is very small, with $\Delta U$ reaching a maximum magnitude of $O(10^{-2})$. The non-monotonicity in $\omega$ observed in b) is a result of the competition between localisation of the response and increase in compressive strain due to increasing $\omega$.

\subsection{Applied Stress}\label{App: Param var stiff - AS}

\begin{figure}[h!]
    \centering
\includegraphics[width=9cm]{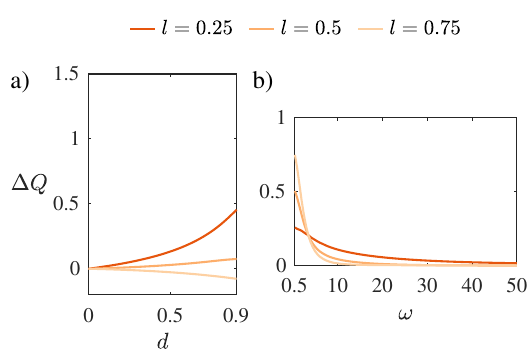}
    \caption{ Difference between between damaged and undamaged net flux against a) $d$ and b) $\omega$ under AS ($A_l=0.2$) for three locations $l$,}
    \label{Fig: Effect params AS Appendix}
\end{figure}

Figure \ref{Fig: Effect params AS Appendix} displays the effect of damage magnitude $d$ and loading frequency $\omega$ on net flux under AS. Net flux increases with $d$ for $l=0.25$ and $l=0.5$ similarly to under AD. For $l=0.75$ however net flux decreases with $d$. Net flux also decreases with $\omega$, though comparatively less than under AD. In addition for $\omega<3.5$, $\Delta Q$ increases with $l$ and for $\omega>3.5$ the relationship is reversed so that $\Delta Q$ decreases with $l$. Decreasing frequency homogenises the response so that maximum strain at $Z=l$ is the same for all $l$; the same increase in strain but farther away from the boundary then requires a greater flux, so that $\Delta Q$ increases with $l$ for small $\omega$.

\subsection{Summary table}\label{App: Param var stiff - table}

We present a table summarising results from our parameter variation, including the sign of $\Delta U$ and $\Delta Q$, their relationship with $d$, $l$ and $\omega$ and their order of magnitude. Upward (downward) arrows are shorthand for increase (decrease), and refer to the absolute value of $\Delta U,\Delta Q$.

\newcolumntype{C}[1]{>{\centering\let\newline\\\arraybackslash\hspace{0pt}}m{#1}}

\begin{center}
 \begin{tabular}{|C{0.5cm}||C{5cm}|C{5cm}|}
    \hline
    & Applied Stress & Applied Displacement \\
    \hline
    \multirow{5}{0.5cm}{$\Delta U$}  & $>0$ & \multirow{5}{1cm}{$\approx 0$} \\
    & $\nearrow$ with $d$ & \\
    & $\approx$ for all $l$ & \\
    & $\searrow$ with $\omega$ & \\
    & $O(10^{-1}-1)$ & \\
    \hline
    \multirow{5}{0.5cm}{$\Delta Q$}  & $>0$ & $>0$ \\
    & $\nearrow$ with $d$ & $\nearrow$ with $d$ \\
    & $\begin{cases} \nearrow \;\text{with}\; l & \omega <3.5 \\ \searrow \;\text{with}\; l & \omega >3.5 \end{cases}$ & $\begin{cases} \nearrow \;\text{with}\; l &d <0.5 \\ \searrow \;\text{with}\; l & d >0.5 \end{cases}$ \\
    & $\searrow$ with $\omega$ & $\searrow$ with $\omega$\\
    & $O(10^{-1})$ & $O(10^{-1}-10)$\\
    \hline
\end{tabular}
\end{center}

\section{Permeability damage}\label{App: perm}

\subsection{Full solution}\label{App: perm - full solution}

\begin{figure}
\centering
\includegraphics[width=14.9cm]{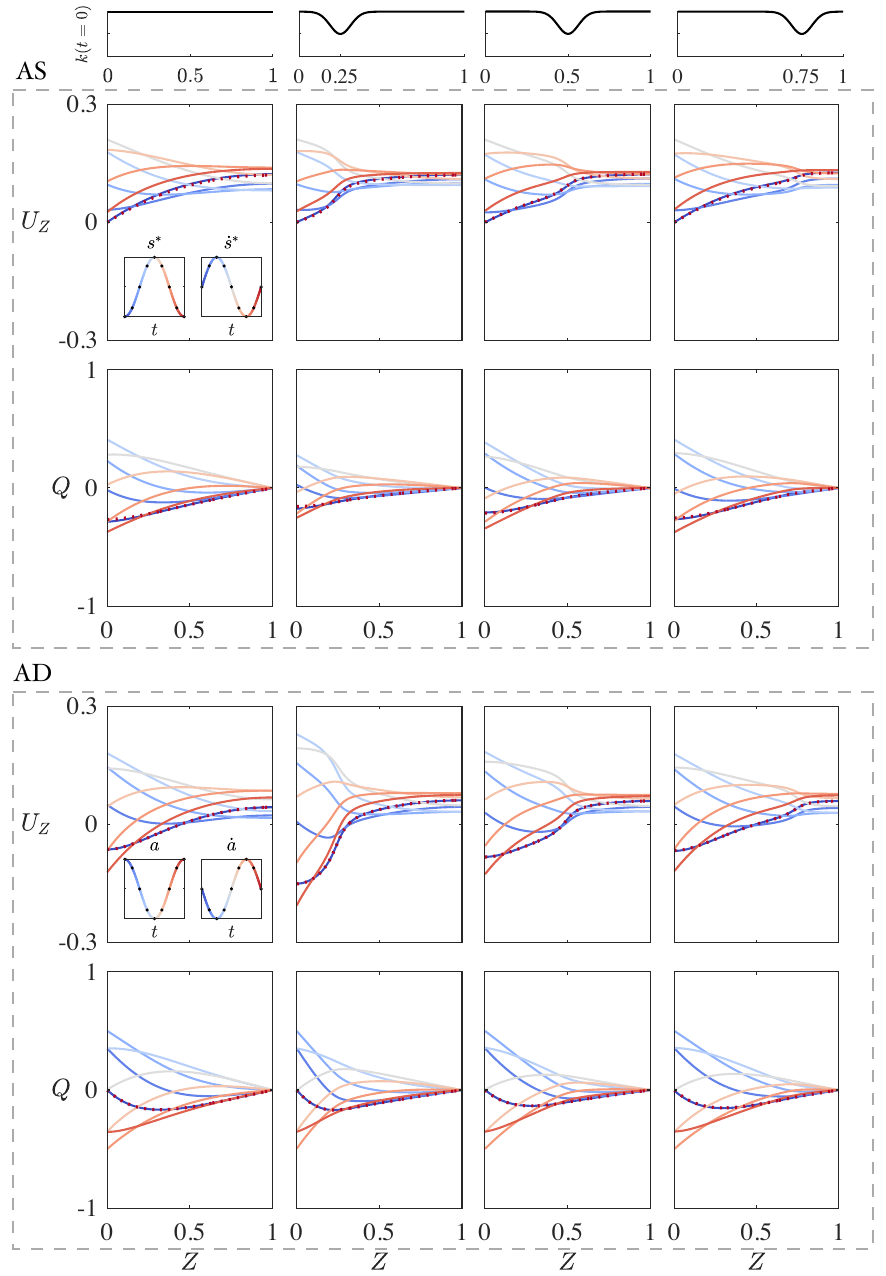}
    \caption{Strain (first and third rows) and flux (second and fourth rows) plotted for 9 equally spaced values of cycle $38\pi/\omega \leq t < 40\pi/\omega$, for uniform stiffness (first column) and locally decreased permeability with $d = 0.8$(columns 2-4). The first two rows are in response to an applied stress ($A_l=0.2$, $\omega=10$) and the second two rows are in response to an applied displacement ($A_d=0.1$, $\omega=10$). The applied stress $s^*(t)$ and displacement $A(t)$, and their respective time derivatives $\Dot{s}^*(t)$, $\Dot{a}(t)$ are displayed as insets in the first column. We differentiate between the loading phase ($\Dot{s}^*>0$, $\Dot{a}<0$, dark blue to light blue) and unloading phase ($\Dot{s}^*<0$,$\Dot{a}>0$, light red to dark red).}
    \label{Fig: Strain and flux profiles with perm damage}
\end{figure}

Figure \ref{Fig: Strain and flux profiles with perm damage} displays the full strain and flux solutions over one cycle for 9 equally spaced values of $38\pi/\omega\leq t_i< 40\pi/\omega$, under an applied stress (AS) with $(A_l,\omega)=(0.2,10)$ in the first two rows and under an applied displacement (AD) with $(A_d,\omega)=(0.1,10)$ in the last two rows. We differentiate between the loading phase, $\Dot{s}^*>0$ or $\Dot{a}<0$ (dark blue to light blue) and unloading phase, $\Dot{s}^*<0$ or $\Dot{a}>0$ (light red to dark red).  The first column is the homogeneous case for reference and columns 2-4 display profiles with imposed local decrease in permeability with $d = 0.8$ where the location of the dip is increased from left to right (moving away from the moving boundary). 

On the one hand under AS there is a clear decrease in strain for $Z>l$ (first row), since diffusion is decreased. This is also accompanied with a decrease in flux (second row). Under AD on the other hand, strain is increased for $Z<l$ (third row) and flux decays more rapidly with $Z$ for $Z>l$ (fourth row). We note that these results are for $d=0.8$ whilst the profiles for $d=0.35$ show very little change compared to the homogeneous case.

\subsection{Summary table of parameter variation}\label{App: perm - table}

We present a table summarising results from our parameter variation, including the sign of $\Delta U$ and $\Delta Q$, their relationship with $d$, $l$ and $\omega$ and their order of magnitude. Upward (downward) arrows are shorthand for increase (decrease), and refer to the absolute value of $\Delta U,\Delta Q$.

\begin{center}
 \begin{tabular}{|C{0.5cm}||C{5cm}|C{5cm}|}
    \hline
    & Applied Stress & Applied Displacement \\
    \hline
    \multirow{5}{0.5cm}{$\Delta U$}  & $\begin{cases} >0 & \omega <5.5 \\ <0 & \omega >5.5 \end{cases}$ & $>0$ \\
    & $\nearrow$ with $d$ & $\nearrow$ with $d$\\
    & $\searrow$ with $l$ & $\searrow$ with $l$ \\
    & $\begin{cases} \nearrow \;\text{with}\; \omega & \omega <1.5 \\ \searrow \;\text{with}\; \omega & \omega >1.5 \end{cases}$ & $\begin{cases} \nearrow \;\text{with}\; \omega & \omega <5.5 \\ \searrow \;\text{with}\; \omega & \omega >5.5 \end{cases}$\\
    & $O(10^{-2}-10^{-1})$ & $O(10^{-2}-10^{-1})$\\
    \hline
    \multirow{5}{0.5cm}{$\Delta Q$}  & $<0$ & $<0$ \\
    & $\nearrow$ with $d$ & $\nearrow$ with $d$ \\
    & $\searrow$ with $l$ & $\searrow$ with $l$ \\
    & $\begin{cases} \nearrow \;\text{with}\; \omega & \omega <4 \\ \searrow \;\text{with}\; \omega & \omega >4 \end{cases}$ & $\begin{cases} \nearrow \;\text{with}\; \omega & \omega <11 \\ \searrow \;\text{with}\; \omega & \omega >11 \end{cases}$\\
    & $O(10^{-1})$ & $O(10^{-1})$\\
    \hline
\end{tabular}
\end{center}

\bibliographystyle{IEEEtran}
\bibliography{Paper}

\end{document}